\documentclass[preprint,12pt]{elsarticle}
\usepackage{amsmath}
\usepackage{amssymb}
\usepackage{graphicx}
\usepackage{epstopdf, epsfig}
\usepackage{subfig,float}
\usepackage{natbib}
\usepackage{xcolor}
\usepackage{booktabs}
\usepackage{url}
\usepackage{enumitem}
\usepackage[percent]{overpic}

\definecolor{orange}{rgb}{1.0,0.4,0.0}

\newcommand{\bds}[1]{\boldsymbol{#1}}

\newcommand{\etal}{{\textit{et al. }}}

\DeclareFontFamily{OT1}{pzc}{}
\DeclareFontShape{OT1}{pzc}{m}{it}{<-> s * [1.10] pzcmi7t}{}
\DeclareMathAlphabet{\mathpzc}{OT1}{pzc}{m}{it}

\def\etal{et\penalty50\ al.~}

\definecolor{RefOneColor}{RGB}{0,70,140}

\definecolor{RefTwoColor}{RGB}{150,40,40}

\definecolor{MajorColor}{RGB}{0,120,60}

\journal{Computers and Fluids (accepted version)}

\begin{document}
\title{Analysis and simulations of droplet generation regimes in a coaxial microfluidic device}
\author{Alessio Innocenti\corref{cor1}}
\ead{alessio.innocenti@unipi.it}
\cortext[cor1]{Corresponding author}

\author{Andrea Poggi}

\author{Simone Camarri}

\author{Maria Vittoria Salvetti}

\affiliation{organization={University of Pisa, Dipartimento di Ingegneria Civile e Industriale},
addressline={Largo Lucio Lazzarino},
postcode={56122},
city={Pisa},
country={Italy}}

\begin{abstract}
The generation of microdroplets via segmentation in microfluidic devices is of interest in many applications {ranging from biochemistry to pharmaceuticals}.
This technique permits indeed much higher control on the droplet size, uniformity and generation rate than in standard batch generation processes. 
In this work, we {assess} the suitability of the open-source software Basilisk to accurately predict microdroplet generation by segmentation. We have validated the numerical tool {against} analytical solutions for the dynamics of droplets in confined flows, namely with Bretherton's theory, and by comparison with experimental results {available in the literature}. We have then performed {several numerical simulations} for a coaxial device, analyzing the different regimes of droplet generation, and evaluating how the physical and flow parameters affect the production mechanisms and {the diameters of the generated droplets}. {Finally, we have proposed new scaling laws for predicting droplet diameters in the dripping and jetting regimes, extending existing correlations by accounting for additional physical effects, such as the viscosity ratio.}
\end{abstract}

\maketitle

\section{Introduction}
{The processes involved in droplet generation at the milli- to micro-scale have attracted considerable attention due to the numerous applications involving microdroplets. These applications include:

\noindent (i) Lab-on-a-chip devices, which are miniaturized platforms reproducing laboratory processes at the micro/milli-scale and enabling sample preparation, mixing, separation, and detection.

\noindent (ii) Drug delivery systems based on microparticles, nanoparticles, coated microparticles, and microencapsulation. These systems offer the significant advantage of delivering small drug doses selectively to the target site, thereby reducing accumulation and undesired systemic side effects. Moreover, they provide improved control over drug release rates through modifications of the surrounding microenvironment \cite{vargason2021evolution}.

\noindent (iii) Encapsulation and manipulation of single cells within aqueous droplets for cell culture applications. This approach allows each individual cell to be isolated within a controlled microenvironment, enabling detailed investigations of cellular behavior and cell–cell interactions \cite{koster2008drop}.} 

{Compared with batch methods, droplet generation by segmentation in microfluidic devices offers excellent control over both droplet size distribution and generation rate, enabling the production of nearly monodisperse droplets. These advantages are achieved at the expense of lower throughput.} The generation of microdroplets via segmentation is generally achieved at a junction of separate channels fed with immiscible fluids, {typically} oil and an aqueous solution, through different possible regimes, namely squeezing, dripping and jetting \cite{anna2016droplets}. The analysis of different design configurations, geometries and {responses} to variations of physical parameters is crucial {for optimizing these devices and maintaining control over the production regime, droplet volume, production rate, and monodispersity.} The geometrical configurations can range from cross-flow and T-junction, to co-flow, and flow-focusing setups. {The} forces driving the breakup instability are surface tension, which has a stabilizing effect, on one side, and viscous stresses from the outer fluid and inertial forces of the inner fluid, which have destabilizing effects, on the other side. {The balance of these forces} can be quantified through the dimensionless capillary and Weber numbers. At very low capillary number the stabilizing effect of surface tension {prevails over} viscous stresses. Therefore at the junction the jet tip of the dispersed phase grows until it fully obstructs the junction, causing a sudden increase in the upstream pressure {that leads to necking and jet breakup.} {At larger capillary numbers, the system can reach the marginal equilibrium of forces, {resulting in }the dripping regime, or it can undergo a convective instability driven by the Rayleigh–Plateau instability of the jet interface, leading to the jetting regime.} The different regimes of formation and the underlying physics have been {extensively} studied in experimental literature {\cite{fu2012droplet, garstecki2005mechanism, cubaud2008capillary, wankawala2023experimental, wankawala2025experiments, wankawala2025enhanced}}, and more recently also with numerical approaches, which have the advantages of more flexibility of use, lower costs and easier inspection of physical fields. 
{However, even if the fluid dynamics of such devices may appear  simple at first glance, the presence of fluid interfaces can significantly complicate the numerical solution. While the very low Reynolds numbers involved imply that the flows are inherently laminar, an inadequate representation of the interface can compromise the accuracy of the solution and, in the worst case, can lead to numerical instabilities due to the amplification of capillary waves.}
This latter complication, with an explicit treatment of surface tension, can only be overcome at the expense of using very small time steps, which greatly increases computational costs. Even if not always all these aspect are considered rigorously, the literature on numerical approaches to droplet generation in microdevices is rapidly increasing, and different methods and softwares are now commonly used in device design. Hua \textit{et al.} \cite{hua2007numerical} simulated coflowing devices with a front-tracking method \cite{unverdi1992front} to track the interface between the two immiscible phases, exploiting mainly the dripping and jetting regimes at different values of the {capillary}  and Weber numbers. De Menech \textit{et al.} \cite{de2008transition} analyzed the transition from squeezing to dripping in a T-junction {using} the phase-field method, which adopts a diffuse interface approach {motivated by the low capillary numbers typically encountered in microfluidic flows, where surface tension effects dominate. They related the breakup dynamics to upstream pressure variations.} Wu \textit{et al.} \cite{wu2008three} performed three dimensional simulations with the lattice Boltzmann method for a cross-junction configuration, with a surface tension calculation based on the continuum surface force (CSF) method, with a good comparison with experiments.
Afkhami and coworkers explored the use of Volume-of-Fluid (VOF) methods for the interface tracking, together with the geometrical height function method for the calculation of curvature and the estimation of surface tension, in a T-junction \cite{afkhami2011numerical} and in a co-flowing configuration in a Hele-Shaw cell \cite{afkhami2013volume}. A different VOF implementation has been adopted in another work \cite{yu2019droplet} to study the effect of the injection angle in a 3-dimensional cross-flow geometry. {The} most widely used numerical tool for such kind of flows probably remains OpenFOAM with its \textit{interFoam} solver, which is {also} based on the VOF method, but with a diffuse interface approach \cite{hoang2012modeling,hoang2013benchmark,fatehifar2021non}. {It is worth mentioning that, in parallel, research is also being conducted using commercial codes, such as COMSOL Multiphysics with a level-set approach \cite{wankawala2025experiments} or ANSYS Fluent, where a VOF implementation is present as well \cite{masoni2024formation}.}

The literature on numerical approaches for microfluidics is therefore highly heterogeneous, with no single method having emerged as the preferred choice over the others. In this work, we aim to investigate the suitability of the open-source software Basilisk (\url{https://basilisk.fr}) for addressing this class of applications. Basilisk is extensively used for the simulation of multiphase flows and droplet dynamics, however, its application to droplet generation processes in microfluidic devices has received limited attention. 
Recently, Harispe \etal \cite{harispe2024accurate} {used} Basilisk to evaluate its applicability to droplet generation in coaxial and T-junction configurations with encouraging results, although {considering only} a limited range of configurations and parameters. Our intent is, therefore, to investigate deeper some peculiar numerical aspects of Basilisk in the application to droplet generation in microfluidic devices, assessing its accuracy against theoretical and experimental results, and exploiting its usage for droplets regime production and diameter predictions. We then analyze droplet generation regimes in a coaxial device and the influence of physical and flow parameters. New scaling laws are finally proposed to predict droplet diameters in the dripping and jetting regimes, providing an extension of existing correlations by incorporating additional effects such as the viscosity ratio.

The work is structured as follows: in Section \ref{sec:math} we briefly describe the mathematical formulation of the problem; in Section \ref{sec:numerics} we report the details of the numerical implementation of Basilisk; in Section \ref{sec:taylor} we validate and assess the accuracy of Basilisk for a Taylor bubble configuration; in Section \ref{sec:coaxial} we analyze in details a coaxial configuration for droplet generation with validation against literature experiments, regime mapping and derivation of scaling laws for the droplet diameter.


\section{Modeling and governing equations} \label{sec:math}
We describe the dynamics of droplet generation in a microfluidic device, where two {immiscible} fluids enter from separate channels, come into contact at a junction, and flow out {through} the same channel. 
The physical parameters characterizing the problem are the viscosities and densities of the two fluids, $\mu_i$ and $\rho_i$, and the surface tension $\sigma$. The index $i$ denotes the continuous, $c$, or the dispersed, $d$, phase. Once the geometry is fixed (e.g. coaxial, cross junction, etc.), the additional parameters characterizing the flow are the inlet velocities of the two fluids, $U_c$ and $U_d$, and a characteristic dimension of the channels, typically the diameter, $D$, in coaxial configurations or the channel height, $H$, in {rectangular} section channels. Based on these physical, geometrical and flow parameters, four dimensionless groups can be formed: firstly the viscosity and density ratios, respectively $\mu_d/\mu_c$ and $\rho_d / \rho_c$; secondly the Reynolds and capillary {numbers}, {which, for this kind of flows,}  can be defined based on the continuous phase properties as follows
\begin{eqnarray}
Re_c = \frac{\rho_c U_c D}{\mu_c}\; , \qquad Ca_c = \frac{\mu_c U_c}{\sigma}
\end{eqnarray}

An alternative, quite common in microfluidics literature, is to replace one of these two dimensionless {numbers} with the Weber number, defined as follows:
\begin{eqnarray}
We_c = \frac{\rho_c U_c^2 D}{\sigma} = Re_c \cdot Ca_c
\end{eqnarray}

The two fluids are governed by the {incompressible} Navier-Stokes equations with variable density and viscosity, which, {in nondimensional form, are written as follows:}
\begin{eqnarray}
\nabla \cdot \mathbf{u} &=& 0 \\
\varphi_r \Bigl ( \frac{\partial {\bf u}}{\partial t} + {\bf u} \cdot \nabla {\bf u} \Bigl ) &=&  - \nabla p + \frac{1}{Re_c} \nabla \cdot (2 \lambda_r \bds{ D}) + \frac{1}{Re_c} \frac{1}{Ca_c} \kappa {\bf n} \delta_s.
\label{eq:ns}
\end{eqnarray}

 $\bds{ D}= [\nabla u +  (\nabla u)^{\textbf{T}}]/2$ is the symmetric strain-rate; $\varphi_r = \rho({\bf x},t) / \rho_c$ is a dimensionless density; $\lambda_r = \mu ({\bf x},t) / \mu_c$ is a dimensionless viscosity.
 The last term represents a non-dimensional surface tension with $\kappa$ and ${\bf n}$ being the local curvature and the interface unit normal, and $\delta_s$ representing a Dirac delta function that is non-zero only at the interface. We do not consider variable surface tension over the interface separating the two fluids.
 
The volume fraction $C$ is used to distinguish between the different phases, being $C = 1$ in the continuous phase, $C = 0$ in the dispersed one, and $0 < C < 1$ at the interface. This scalar field is advected by the equation:
 \begin{eqnarray}
 \frac{\partial}{\partial t} C + {\bf u} \cdot \nabla C = 0 ,
 \end{eqnarray}
 
 The variable density and viscosity are evaluated {by} weighting the continuous and the dispersed {contributions using} the local volume fraction as follows:
  \begin{eqnarray}
\rho({\bf x},t) = C \rho_c + (1 - C) \rho_d \; , \quad
\mu({\bf x},t) = C \mu_c + (1 - C) \mu_d \; ,
 \end{eqnarray}

This set of equations is numerically solved {using} Basilisk.

\section{Numerical implementation} \label{sec:numerics}
Basilisk (\url{http://basilisk.fr}) is a library of solvers that can be used for a variety of fluid dynamics applications, from industrial to geophysical flows. Basic solvers and functions can be included in a script in order to build a specific solver assembly with the desired features. All the different parts share a common discretization based on the finite-volume method on Cartesian {grids}. The grid can also be dynamically refined or coarsened based on specific criteria and relying on a quadtree (2-d) or octree (3-d) refinement strategy \cite{popinet2009accurate}. {In this work, we make use of the centered Navier–Stokes formulation, where the spatial discretization is performed using a centered finite-volume method, while the temporal discretization is carried out through a second-order fractional-step method with a staggered-in-time discretization of the velocity and scalar fields \cite{popinet2009accurate}. The advection term is estimated using the second-order Bell–Colella–Glaz unsplit upwind scheme \cite{popinet2003gerris}.} The Poisson equation is solved iteratively using an efficient multilevel solver constructed on the quad/octree grid. More details can be found at \url{http://basilisk.fr}. All simulations performed {herein} with Basilisk have been done with adaptive mesh refinement (AMR) on quadtree grids.

The interface is advected {using} a geometric piecewise-linear interface construction (PLIC) VOF method (\cite{hirt1981volume,tryggvason2011direct}), generalized for the quad/octree spatial discretization, which can ensure a sharp interface while preserving mass conservation. The scheme is made of two steps: (i) first, the interface is geometrically reconstructed, (ii) then, the fluxes are evaluated and {the} interface is advected. In the PLIC method the interface is approximated by a local segment of equation ${\bf m} \cdot {\bf x} = \alpha$. The normal m is determined {from} the volume fraction in the cell and in the neighbouring ones using the Mixed-Youngs-Centered scheme \cite{aulisa2007interface}. Then, the interface line is moved along the normal by changing $\alpha$ to obtain the desired volume fraction. 
{The advection equation is then integrated using a dimension-splitting approach, which solves the equation successively along each spatial dimension through a fractional-step method that guarantees exact mass conservation.}

The surface tension term in Equation (\ref{eq:ns}) is modeled {through} the continuum surface force (CSF) approach \cite{brackbill1992continuum}. Accurate computation of the curvature is carried out using the height-function method \cite{popinet2009accurate}.

The device geometry is defined in Basilisk through the embedded boundary method (EBM), originally proposed in \cite{johansen1998cartesian,schwartz2006cartesian}, and widely used in {the Basilisk-related} literature for moving boundaries \cite{ghigo2021conservative}, solidification modeling \cite{limare2023hybrid}, and two-phase flows {in} complex geometries \cite{tavares2024coupled}. The method embeds a sharp representation of a solid surface on a regular Cartesian grid. The reconstruction of the solid interface is {obtained through} a piecewise linear reconstruction, using a signed distance function sampled on the vertices of the Cartesian grid for the geometrical properties. The discrete operators in the cut cells are then properly defined with the finite volume method with 2nd order accuracy \cite{tavares2024coupled} for single-phase flow, {and} 1st order accuracy when coupled with the VOF method.

Concerning the boundary condition of the fluid volume fraction at the cut cell, a particular treatment must be adopted when in the same cell there are the two liquid phases and the solid, indeed a condition on the contact angle {needs} to be set \cite{afkhami2009mesh,legendre2015comparison}. For more details on the numerical implementation of the contact angle condition in Basilisk, see \cite{tavares2024coupled}. In all the present simulations, when the contact between the interface and the wall is present, we have used an equilibrium contact angle of $\alpha_w = 30^{\circ}$ (measured between the wall and the interface in the continuous phase), {which is within the typical range of values adopted in numerical simulations \cite{masoni2024formation, fatehifar2021non, hoang2013benchmark}}. {A precise tuning of the value of the equilibrium contact angle, which depends on the specific properties of the fluid and of the solid surface, was not carried out, as it would require a more detailed characterization of the experimental conditions. {This} is beyond the scope of the present work, as the comparison is performed with an experiment from the literature.}

For the velocity and pressure fields we set an inlet boundary condition with constant velocity at the entrance of the channel, a no-slip condition at the walls, and an outflow condition with fixed pressure at the outlet of the channel. In the axially symmetric configuration {symmetry conditions are set} on the symmetry axis, which is located at the bottom of the domain ($y=0$). More details on geometries and boundary conditions are given in the following sections.

\section{{Two}-dimensional Taylor bubble} \label{sec:taylor}
Before studying the droplet generation in a coaxial device, we assess the accuracy of Basilisk in reproducing the dynamics of a Taylor bubble, since the literature on Basilisk applications to problems with bubbles/droplets in microchannels is still very limited to date. In particular we analyze a 2-dimensional configuration with an already formed droplet/bubble in a straight channel at small values of $Re$ and $Ca$ numbers. We try to reproduce the results of Fullana \textit{et al.} \cite{fullana2015droplet}, a precursor work focused on {the} Gerris code ({the predecessor} of Basilisk). The geometry consists therefore {of} a straight channel of height $H$ with an already formed droplet of initial ellipsoidal shape, with equivalent diameter larger than the channel height. A constant inlet velocity is used as boundary condition on one side of the channel, while an outflow boundary condition is used on the other side, with no-slip at the walls. After a transient where the droplet deforms and accelerates, a steady shape and velocity are reached within {a few fractions of the} diameter downstream of the initial position. The Reynolds number is kept fixed with a value of $Re = 0.1$, while the {capillary} number is varied in the range $Ca = [0.0005 - 0.1]$. The density and viscosity ratios are kept constant {equal to one}, i.e. $\rho_d / \rho_c = 1$, $\mu_d / \mu_c = 1$.  

\begin{figure}[h]
\center
{\includegraphics[width=.48\textwidth]{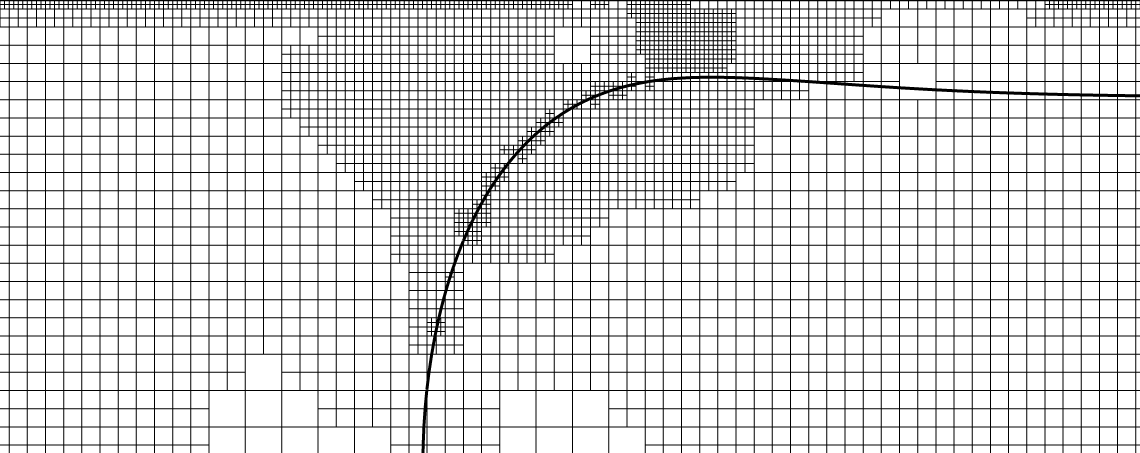}}
 {\includegraphics[width=.48\textwidth]{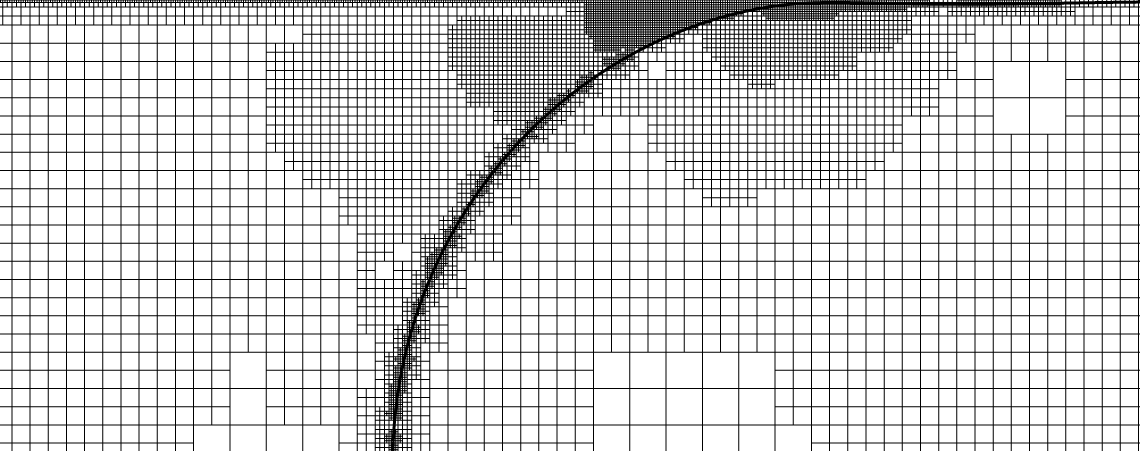}}
\caption{Example of grid adaptivity for the Taylor bubble at $Re=0.1$. Left panel $Ca = 0.1$, maximum resolution $H/\Delta = 204$, right panel $Ca = 0.001$, maximum resolution $H/\Delta = 410$.}
\label{Fig:droplet_grid}
\end{figure}

{The criteria for adaptive grid refinement} have been set on the two components of velocity, solid fraction and VOF volume fraction, {with the respective thresholds defined as follows:} $\epsilon_{u_x} = \epsilon_{u_y} = 0.007$, $\epsilon_{cs} = 10^{-3}$, $\epsilon_{f} = 10^{-1}$. Indeed, as can be seen in Figure \ref{Fig:droplet_grid}, which shows an example of grid resolution and adaptivity for two different cases of {capillary} numbers, the refinement is effective at the interface, at the solid wall, and at the rear of the droplet where the distance between the interface and the wall is smaller and the gap velocity is therefore higher.

We compare our results with the analytical solution of {Bretherton's} theory \cite{bretherton1961motion,fullana2015droplet}, which gives the film thickness, $t$, between the droplet and the wall, as a function of the {capillary} number as follows:
\begin{equation}
\frac{2 t}{H} = 0.643 (3 Ca)^{2/3}  \; ,
\label{Eq:bretherton}
\end{equation}
and with {the semi-empirical relation of Aussillous \etal \cite{aussillous2000quick}, which goes beyond the limit of very small {capillary} number, and yields:}
\begin{equation}
\frac{2 t}{H} = \frac{0.643 (3 Ca)^{2/3}}{1 + 0.643 \cdot 2.50 \cdot (3Ca)^{2/3}} \;.
\label{Eq:aussillous}
\end{equation}
From both relations we can then derive the droplet dimensionless velocity, applying those results to a Poiseuille flow \cite{hoang2013benchmark}:
\begin{equation}
\frac{U_d}{U_f} = \frac{1}{1 - 2 t / H} \; ,
\label{Eq:poiseuille}
\end{equation}

We evaluate the droplet velocity by {volume-averaging} the velocity field over the {entire droplet volume} and we {then }compare the average droplet velocity with Equation \ref{Eq:poiseuille}.

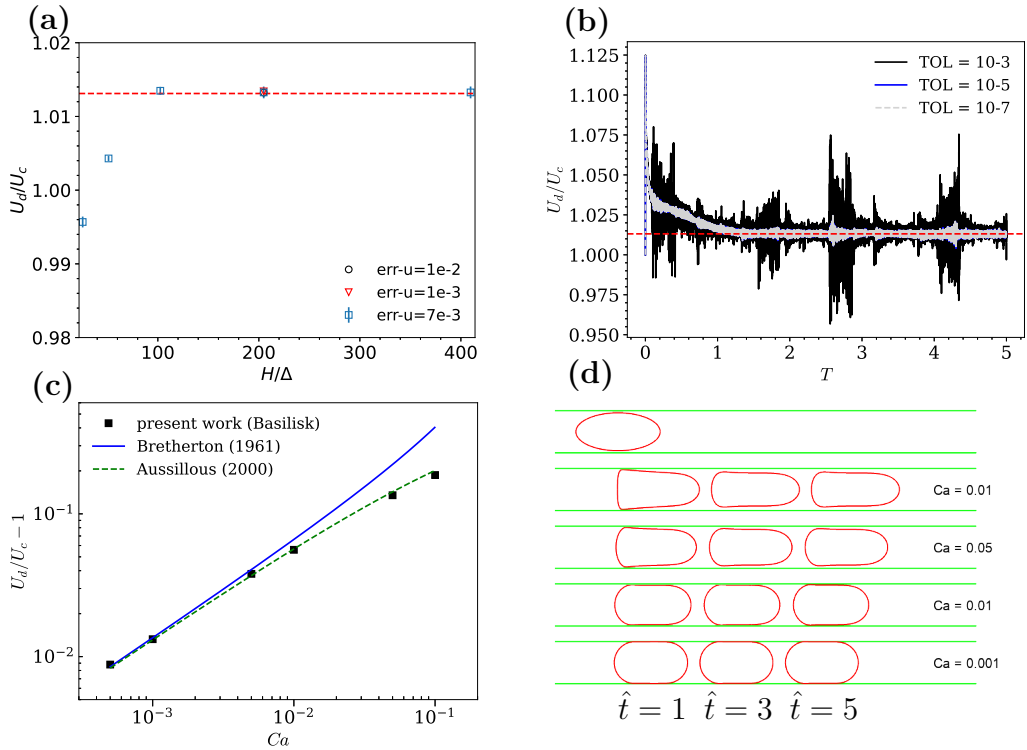
\begin{figure}[ht]
\centering
\begin{minipage}[t]{0.47\textwidth}
	\centering
	\begin{overpic}[width=\linewidth]{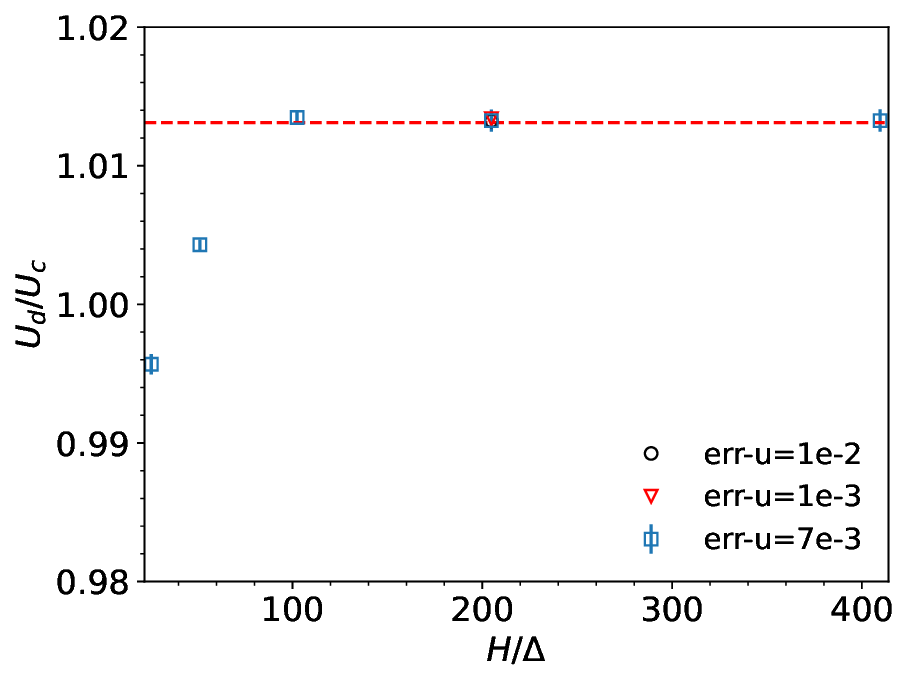}
  		\put(5,75){\textbf{(a)}}
	\end{overpic}
\end{minipage}\hfill
\begin{minipage}[t]{0.48\textwidth}
	\centering
	\begin{overpic}[width=\linewidth]{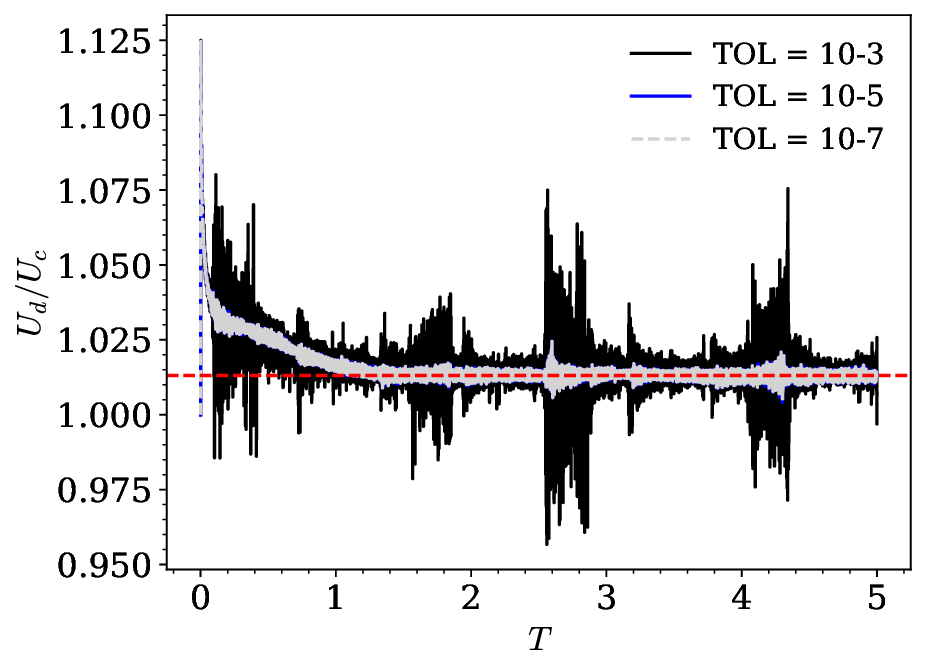}
		\put(5,73){\textbf{(b)}}
	\end{overpic}
\end{minipage}\hfill \\
\begin{minipage}[t]{0.47\textwidth}
	\centering
	\begin{overpic}[width=\linewidth]{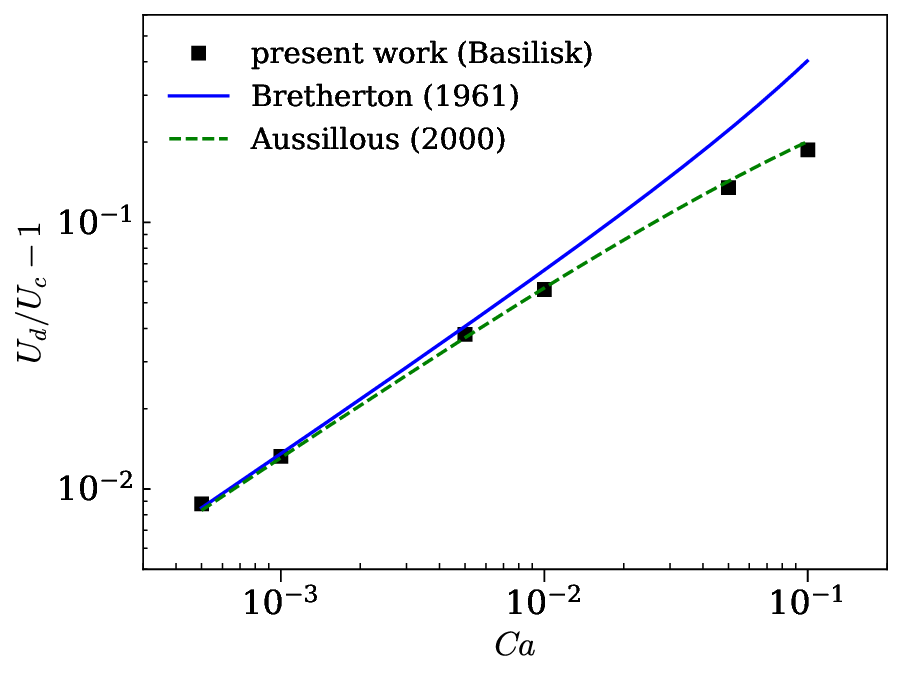}
		\put(6,75){\textbf{(c)}}
	\end{overpic}
\end{minipage}\hfill	
\begin{minipage}[t]{0.52\textwidth}
	\centering
	\vspace{-4.7cm}
	\begin{overpic}[width=1\textwidth]{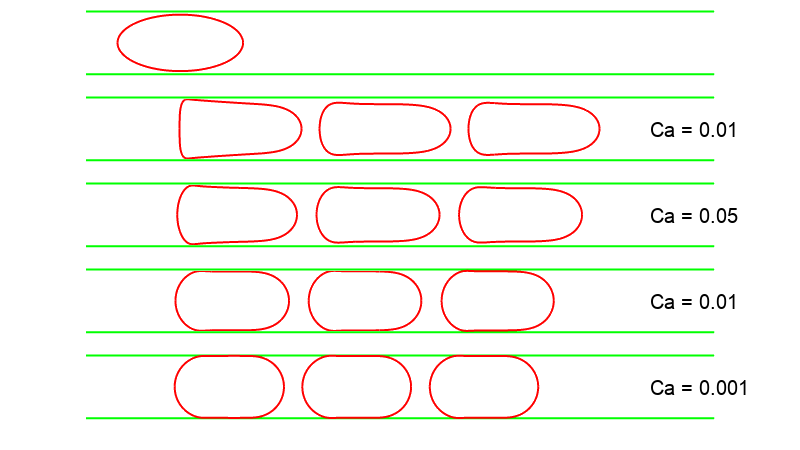}
    		\put(29,0){\makebox(0,0){$\hat{t} = 1$}} 
		\put(45,0){\makebox(0,0){$\hat{t} = 3$}}
		\put(61,0){\makebox(0,0){$\hat{t} = 5$}}
		\put(13,60){\textbf{(d)}}
	\end{overpic}
\end{minipage}
\caption{(a) Grid convergence for $Ca=0.001$ and $Re=0.1$: droplet terminal velocity at different grid {resolutions} and for different adaptivity thresholds. {The red line represents the steady state value obtained {using} the semi-empirical relations \cite{aussillous2000quick}.} (b) Temporal evolution of the droplet terminal velocity with different {tolerances} for the multigrid {resolution of the }Poisson equation. (c) Droplet terminal velocity at different {capillary} numbers compared to {Bretherton's} theory \cite{bretherton1961motion} and literature semi-empirical relations \cite{aussillous2000quick}. (d) {Snapshots of droplet shapes for simulations with different {capillary} numbers.}}
\label{Fig:droplet_scaling_shapes}
\end{figure}

Small {capillary} numbers, as those typical of microfluidics and of this study, {result} in thin gaps between the droplet and the wall, and correctly resolving that region with a sufficient grid resolution is crucial in identifying the droplet shape and in turn its terminal velocity. 
For this reason, we have performed, for the case with $Ca = 0.001$, a series of convergence tests on the most relevant numerical parameters: (i) the maximum level of {grid }refinement; (ii) the thresholds used to dynamically refine/{coarsen} the grid; (iii) the tolerance used in the successive multigrid resolution of the Poisson equation.
The maximum refinement level has been varied in the range $LEV = [8 - 12]$, resulting in a maximum grid resolution in the range $H/\Delta = [25 - 410]$, {with a total number of cells at steady state of $N\simeq 4 \cdot 10^4$ for $H/\Delta \simeq 100$;  $N\simeq 6 \cdot 10^4$ for $H/\Delta \simeq 200$; $N\simeq 9.8 \cdot 10^4$ for $H/\Delta \simeq 400$.} Figure \ref{Fig:droplet_scaling_shapes}(a) shows the droplet velocity with the different grid {resolutions}. It is clear that a resolution between $H/\Delta = 100 - 200$ is necessary, therefore for the other simulations we will adopt a grid resolution of  $H/\Delta \simeq 200$ (except for the smallest $Ca = 0.0005$ for which we have refined up to $H/\Delta \simeq 410$).
We have then analyzed the effect of varying the adaptivity thresholds to refine or {coarsen }the grid. We have found no significant variations {in} the results by varying the criteria for the solid embedded fraction ($cs$) and the volume fraction ($f$), which are not reported here for the sake of brevity. The effect of the adaptivity threshold for velocity has been analyzed in the range $\epsilon_{u_x} = \epsilon_{u_y} = 10^{-2} - 10^{-3}$, showing negligible variations {in} the droplet velocity, as it can be seen in Figure \ref{Fig:droplet_scaling_shapes}(a).

The tolerance parameter of the multigrid {Poisson }solver needs also to be properly tuned to avoid numerical instabilities. Indeed, the very small Reynolds and {capillary}  numbers imply extremely small time {steps}, due to the explicit treatment of surface tension \cite{popinet2009accurate}, and in turn the tolerance of the Poisson equation {solution }might need to be varied with respect to the default value. Figure \ref{Fig:droplet_scaling_shapes}(b) shows the temporal evolution of the droplet velocity with different tolerance values. The large fluctuations observed with too high values are largely reduced with lower values, and from $TOL=10^{-5}$ {the }results are fully converged.

Finally, Figure \ref{Fig:droplet_scaling_shapes}(c) shows the dimensionless droplet velocity for different {capillary}  numbers, compared with Equations \ref{Eq:bretherton}-\ref{Eq:poiseuille}. Results are in excellent agreement with the empirical relation of Aussillous \etal \cite{aussillous2000quick}, especially for the lower {capillary}  numbers, while for $Ca=0.1$ we reach a maximum error of $\sim 6\%$.
{Figure \ref{Fig:droplet_scaling_shapes}(d) shows the contour of droplets for the same range of {capillary}  numbers. Each row represents a simulation with a fixed $Ca$, superimposing successive positions of the same droplet in a single figure. The interfaces were obtained from the VOF field using the geometric PLIC reconstruction provided by Basilisk's \textit{draw\_vof} function, which reconstructs the piecewise-linear interface in each cell from the local volume fraction.}
The shape changes significantly from the classic Bretherton shape to a more symmetric flattened {``pancake-like''} shape at very low {capillary}  numbers.

\section{Coaxial droplet generator} \label{sec:coaxial}
The coaxial configuration analyzed is the one described by \cite{lan2015numerical}, in which the inlet duct for the dispersed phase has an inner diameter $d_{in} = 0.16\,mm$ and an outer diameter $d_{out} = 0.31\,mm$, while the inlet duct for the continuous phase has {a} diameter $D = 0.51\,mm$. The total length of the simulation domain is $L = 20 D$, while the inner duct extends up to $L_{in} = 3D$. The initial position of the interface separating the two phases is a straight line at $x = 2.8 D$, perpendicular to the walls of the inner duct.

\begin{figure}[h]
\center
{\includegraphics[width=.95\textwidth]{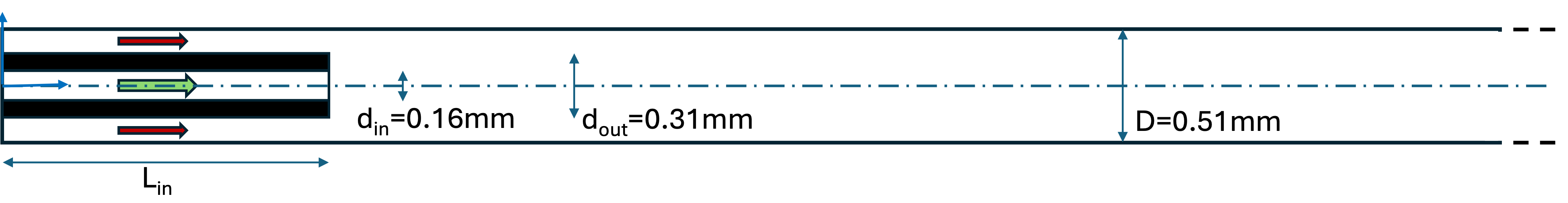}}
\caption{Sketch of the coaxial device.}
\label{Fig:sketch_coaxial}
\end{figure} 

The continuous phase consists of silicone oil of various viscosities (denoted as SO-$\mu$, where $\mu$ is expressed in $[mPa\cdot s]$), while the dispersed phase is a polysulphone solution (PSF) at various concentrations, thus varying the viscosity and slightly the density (which is nevertheless considered constant). Physical properties are reported in Table \ref{Tab:physical_parameters}.

\begin{center}
    \begin{tabular}{|c|c|c|c|c|c|}
    \hline
                                       & SO50  & SO100 & PSF $6\%$  & PSF $8\%$ & PSF $10\%$  \\ \hline
   $\mu [mPa \cdot s]$    & $50$    & $100$  & $9.23$         & $22.7$        & $46.2$           \\
   $\rho [kg/m^{3}]$         & $979$  & $979$  & $969$          & $979$         & $991$            \\
    \hline
    \end{tabular}
    \captionof{table}{Physical parameters.}
	\label{Tab:physical_parameters}
\end{center}

 Before running the complete set of simulations for different combinations of SO and PSF, varying the fluid flow rates to change the dimensionless parameters, we did a preliminary visual validation based on two standard configurations considered in Lan’s article \cite{lan2015numerical}, representing the two main flow regimes: drop flow and jet flow. We set up accordingly the simulations with the same flow rates as in Lan’s work shown in Table \ref{Tab:parameters} below, {and adopted a maximum grid resolution of $D/\Delta = 102$.}

\begin{center}
    \begin{tabular}{|c|c|c|c|c|c|}
    \hline
                    & $\rho_c / \rho_d$ & $\mu_c/\mu_d$ & $\eta = u_d / u_c$ & $Re_c$    & $Ca_c$ \\ \hline
    drop flow & $1$                       & $2.39$          & $1.07$               & $0.0774$  & $0.0824$ \\ \hline
    jet flow    & $1$                       & $2.39$           & $0.915$             & $0.1807$ & $0.1926$  \\
    \hline
    \end{tabular}
    \captionof{table}{Simulation parameters.}
    \label{Tab:parameters}
\end{center}

\begin{figure}[h]
\center
\begin{minipage}[]{0.5\textwidth}
\center
{\includegraphics[width=1.1\textwidth,trim={0 0cm 0 0},clip]{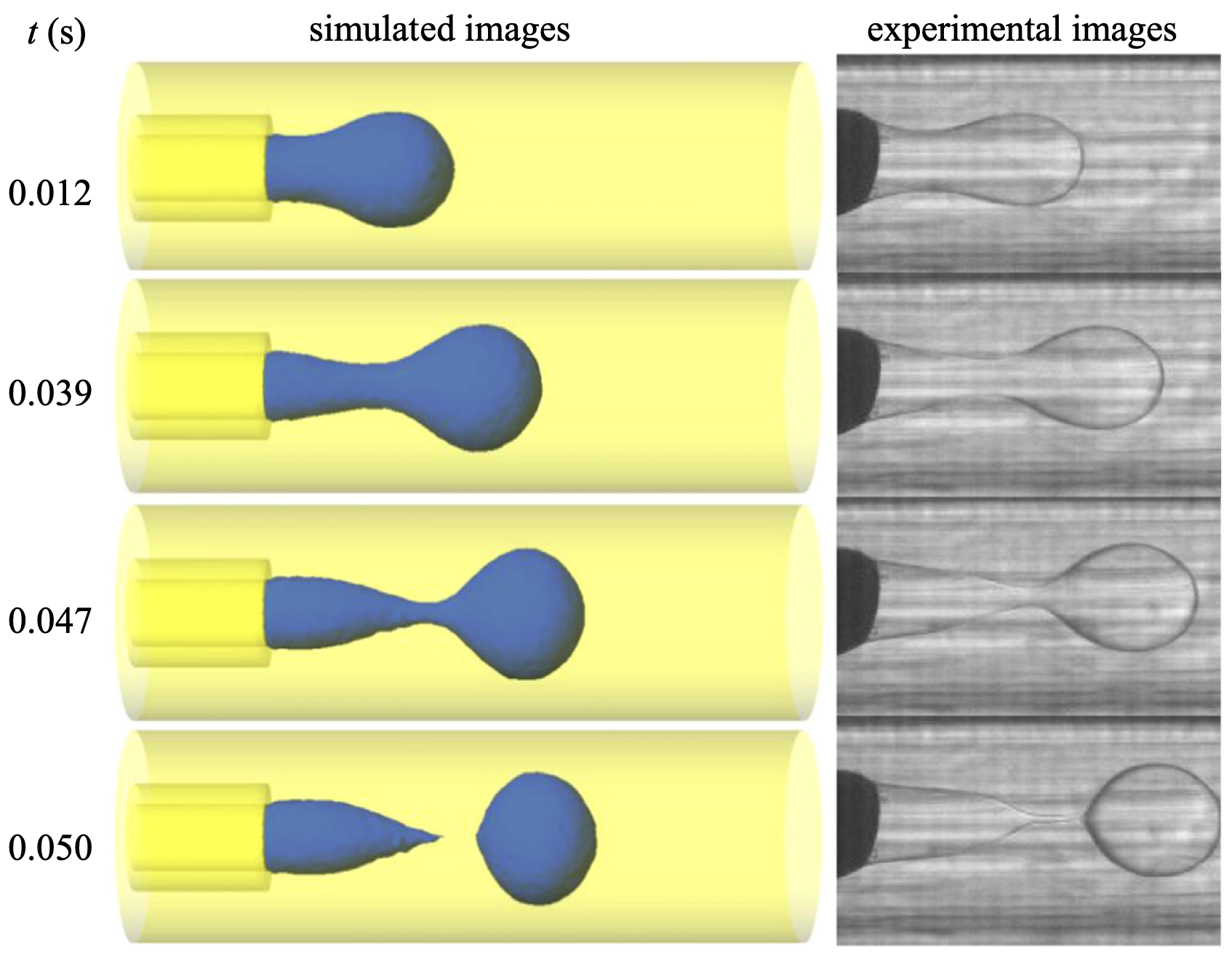}}
\caption*{(a) Simulation and experiments from {the} literature. Reprinted from Chemical Engineering Science, Vol. 134 , W. Lan, S. Li, G. Luo, Numerical and experimental investigation of dripping and jetting flow in a coaxial micro-channel, 76-85 , Copyright 2015, with permission from Elsevier.}
\end{minipage}%
\hfill
\begin{minipage}[]{0.5\textwidth}
\center
\vspace{-2.25cm}
{\includegraphics[width=0.8\textwidth]{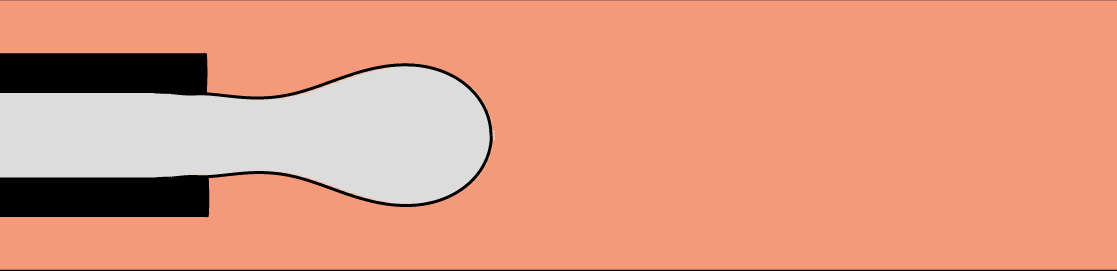}} \\
{\includegraphics[width=0.8\textwidth]{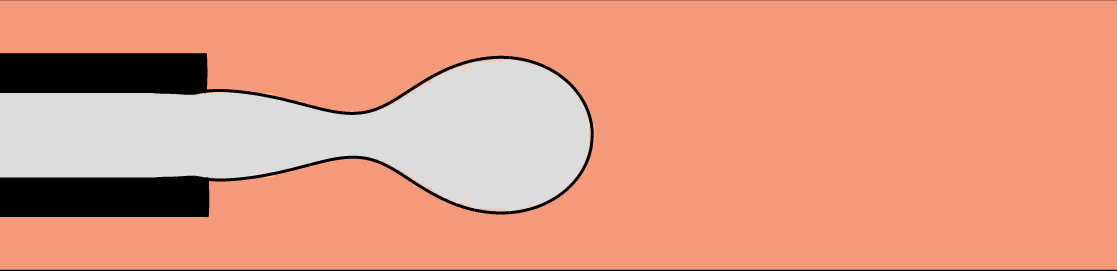}} \\
{\includegraphics[width=0.8\textwidth]{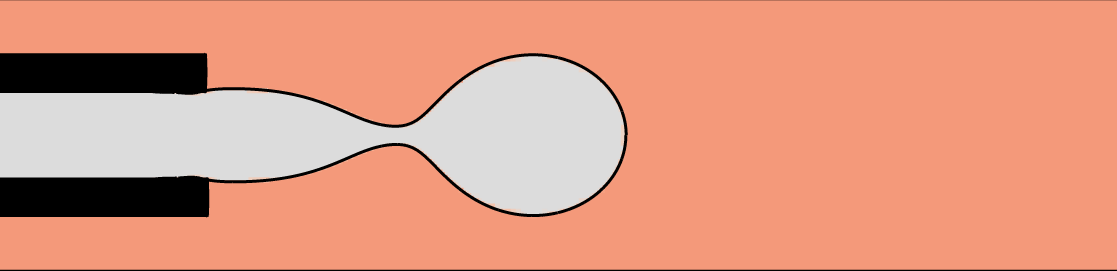}} \\
{\includegraphics[width=0.8\textwidth]{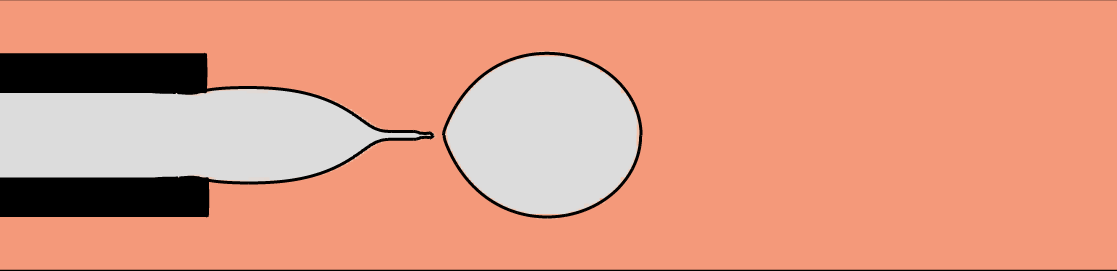}}
\caption*{(b) Basilisk (present work).}
\end{minipage}%
\caption{Droplet formation stages in the dripping regime, comparison with literature results \cite{lan2015numerical}. Basilisk snapshots are at $\hat{t} = 1.72$ ($t = 0.11 s$), $\hat{t} = 2.13$ ($t = 0.14 s$), $\hat{t} = 2.25$ ($t = 0.148 s$), $\hat{t} = 2.30$ ($t = 0.151 s$).}
\label{Fig:experiment_comparison_dripping}
\end{figure}

\begin{figure}[h]
\center
\begin{minipage}[]{0.5\textwidth}
\center
{\includegraphics[width=1.11\textwidth,trim={0 0cm 0 0},clip]{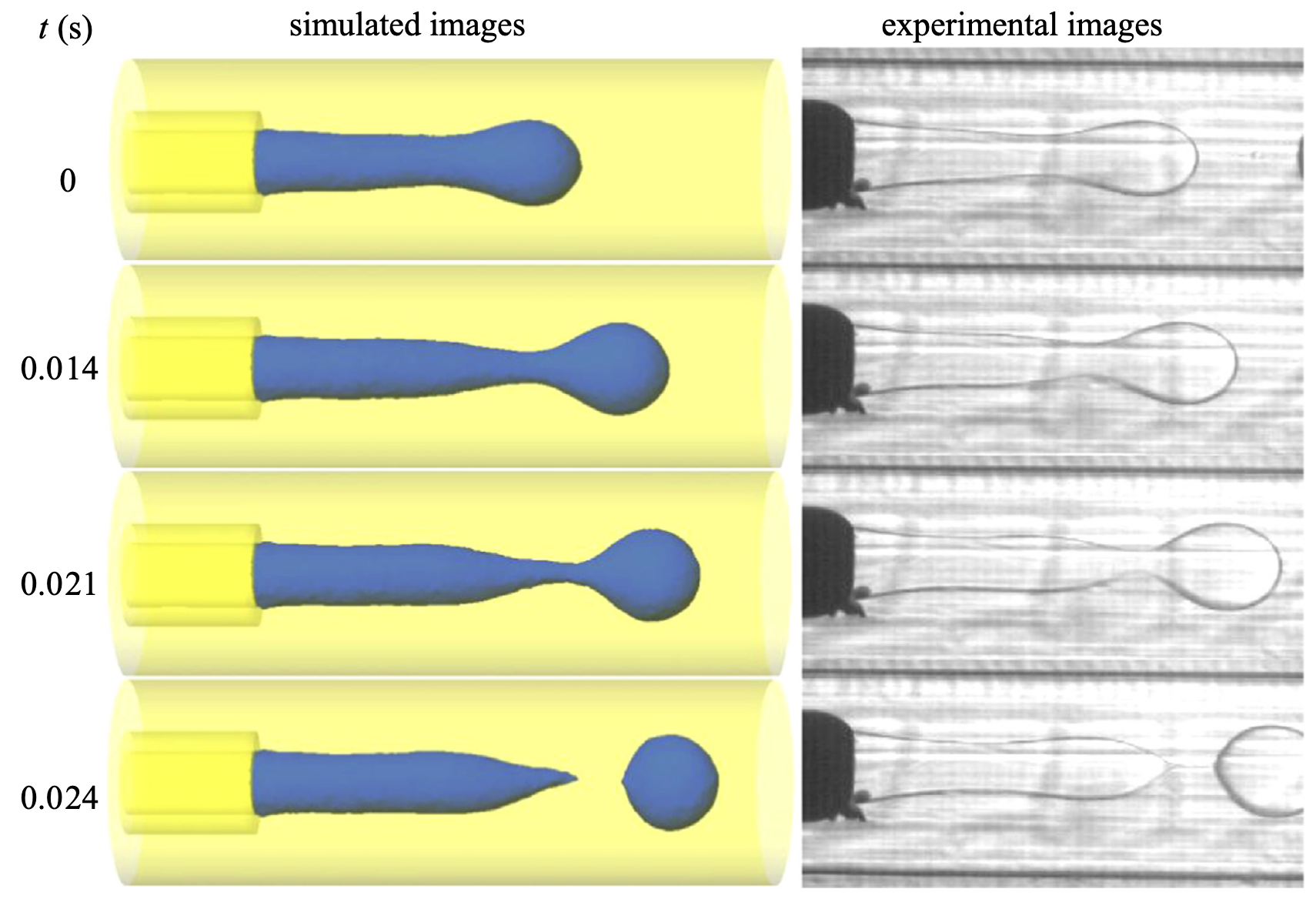}}
\caption*{(a) Simulation and experiments from {the} literature. Reprinted from Chemical Engineering Science, Vol. 134 , W. Lan, S. Li, G. Luo, Numerical and experimental investigation of dripping and jetting flow in a coaxial micro-channel, 76-85 , Copyright 2015, with permission from Elsevier.}
\end{minipage}%
\hfill
\begin{minipage}[]{0.5\textwidth}
\center
\vspace{-2.3cm}
{\includegraphics[width=0.72\textwidth]{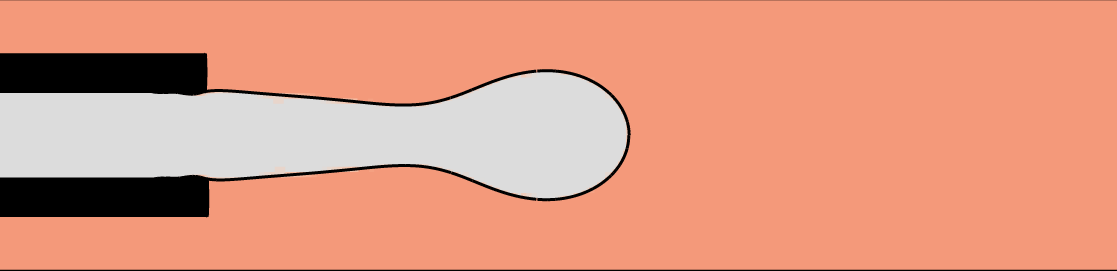}} \\
{\includegraphics[width=0.72\textwidth]{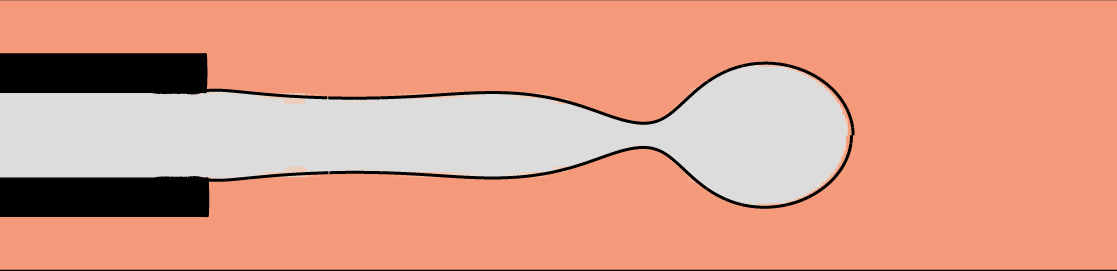}} \\
{\includegraphics[width=0.72\textwidth]{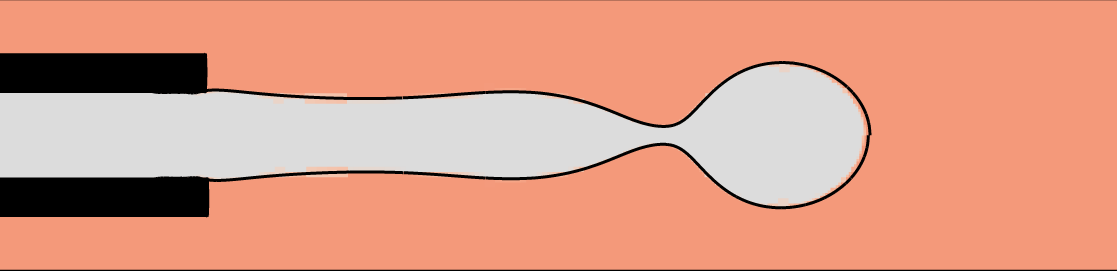}} \\
{\includegraphics[width=0.72\textwidth]{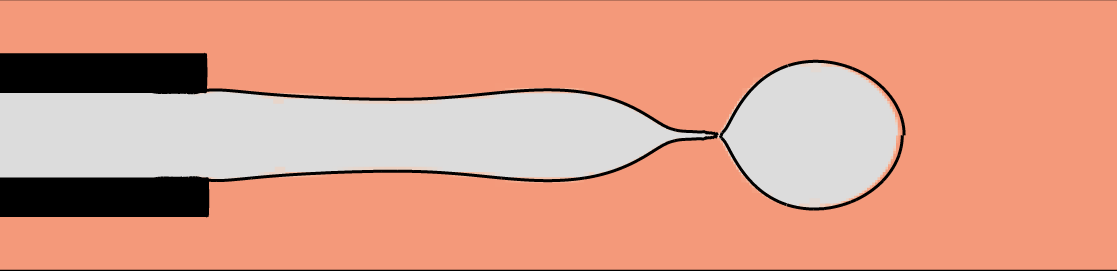}}
\caption*{(b) Basilisk (present work).}
\end{minipage}%
\caption{Droplet formation stages in the jetting regime, comparison with literature results \cite{lan2015numerical}. Basilisk snapshots are at $\hat{t} = 1.75$ ($t = 0.049 s$), $\hat{t} = 2.45$ ($t = 0.063 s$), $\hat{t} = 2.50$ ($t = 0.070 s$), $\hat{t} = 2.60$ ($t = 0.073 s$).}
\label{Fig:experiment_comparison_jetting}
\end{figure}

The comparison {is preformed by} evaluating the results of Basilisk at the same simulation {times as} those of \cite{lan2015numerical}.
Two aspects were considered to reconstruct the images: (i) since the reference zero time of the simulations in \cite{lan2015numerical} {is} not known, we determined comparable instants starting from the droplet detachment snapshot ($t=0.05$) and proceeding backward; (ii) given the geometric symmetries, the simulations in Basilisk were performed considering only the upper half of the channels, and imposing axial symmetry through basilisk $axi.h$ module, which introduces a metric to take into account axial symmetry in a 2D simulation, thus reducing the computational demand. The resulting images therefore have been mirrored with respect to the longitudinal axis to allow a more accurate visual comparison. The visual comparisons are shown in Figure \ref{Fig:experiment_comparison_dripping} for the dripping regime and Figure \ref{Fig:experiment_comparison_jetting} for the jetting regime. 

As {can }be seen the matching of the whole dynamics is good: {the} first important aspect to note is that simulations correctly reproduce the generation mechanisms. Secondly, the diameter and the pinch-off position {are} in agreement. As in the experimental images, at the moment of pinch-off we see the thin tip of the core filament that starts to retract, in both regimes, with an abrupt change of {the tip diameter}. In the jetting regime, we see that there is also a good {correspondence} in terms of Rayleigh-Plateau instability originating the pinch-off, which can be appreciated by undulations of the interface with {a} similar wavelength.

\subsection{Grid Convergence and Adaptivity Criteria} \label{sec:grid_convergence}
{For} the two reference cases shown in the previous section, we have performed a grid convergence analysis with a focus on the refinement parameters. 
We have considered three different levels of maximum refinement, namely $LEV=10,11,12$, which result in $D/\Delta = 51, 102, 204$, with the two sets of refinement parameters reported in Table \ref{Tab:refinement}. 

\begin{center}
    \begin{tabular}{|c|c|c|c|c|}
    \hline
     & $\epsilon_{u_x}$  & $\epsilon_{u_y}$ & $\epsilon_{cs}$  & $\epsilon_{f}$   \\ \hline
   base    & $1.5 \cdot 10^{-2}$                       & $1.5 \cdot 10^{-2}$           & $10^{-7}$             & $10^{-3}$   \\
   ref    & $1.5 \cdot 10^{-3}$                       & $1.5 \cdot 10^{-3}$           & $10^{-7}$             & $10^{-3}$   \\
    \hline
    \end{tabular}
    \captionof{table}{{Sets of threshold values for grid adaptivity.}}
	\label{Tab:refinement}
\end{center}

For each set of refinement parameters, we have carried out a grid convergence study. First of all, we have evaluated the average diameter of the formed droplet and the generation frequency. For the evaluation of droplet diameter we have used {the Basilisk} \textit{tag} function, which allows {us} to identify each closed region of the dispersed phase with a tag, we have calculated the area of each droplet, and the resulting equivalent diameter as:
\begin{equation}
d_{eq} = 2 \sqrt{\frac{2 A_{drop}}{\pi}} \, .
\end{equation}
The average diameter has been calculated by averaging over all droplets present in the domain, starting from the dimensionless time $\hat{t} = 5$. The average generation frequency is evaluated by simply counting the number of {droplets} generated from the first pinch-off up to the dimensionless time $\hat{t} = 20$. Results, reported in Table \ref{Tab:grid_convergence}, show that if the primary interest is on the evaluation of the regime of formation and on the average diameter and production rate, simulations with a maximum level of refinement of $LEV=11$ with the \textit{base} adaptivity set can be considered well converged. $LEV=10$ with the \textit{ref} adaptivity set is fine for the two cases analyzed here, but as it was shown in section § \ref{sec:taylor}, a maximum refinement of $H/\Delta \approx 50$ is not sufficient at small {capillary}  numbers. Therefore $LEV=11$ with the \textit{base} adaptivity set is the best trade-off between accuracy and computational times. 

\begin{center}
    \begin{tabular}{|c|c|c|c|c|}
    \hline
        case            & \multicolumn{2}{|c|}{dripping} & \multicolumn{2}{|c|}{jetting}   \\ \cline{2-5}
     & $D_{drop} / D_{duct}$  & $f_{pinch-off}  [s^{-1}]$ & $D_{drop} / D_{duct}$  & $f_{pinch-off} [s^{-1}]$   \\ \hline
   lev11-base    & $0.6324$                       & $9.23$           & $0.524$             & $32.96$   \\
   lev10-ref    & $0.6362$                       & $9.11$           & $0.527$             & $30.11$   \\
   lev11-ref    & $0.6357$                       & $9.12$           & $0.530$             & $30.11$   \\
   lev12-ref    & $0.6325$                       & $9.37$           & $0.530$             & $30.11$   \\
    \hline
    \end{tabular}
    \captionof{table}{Average diameter and pinch-off frequency of the generated droplets in the two reference simulations. {The left column denotes the combination of maximum level of refinement and adaptivity threshold set.}}
	\label{Tab:grid_convergence}
\end{center}

\begin{figure}[h]
\center
\begin{minipage}[t]{0.48\textwidth}
	\centering
	\begin{overpic}[width=\linewidth]{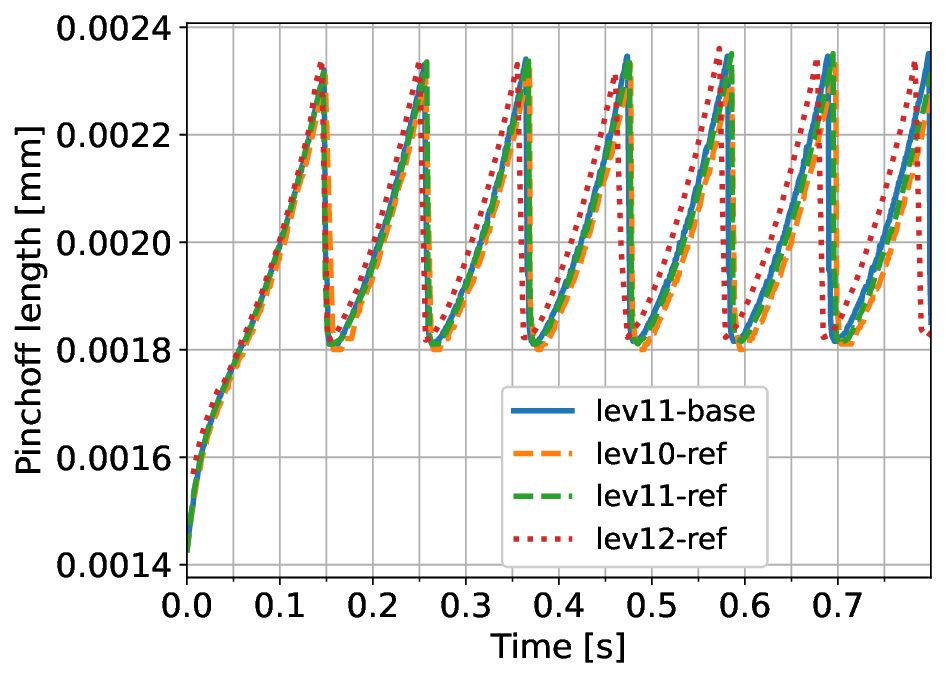}
		\put(5,74){\textbf{(a)}}
	\end{overpic}
\end{minipage}\hfill
\begin{minipage}[t]{0.48\textwidth}
	\centering
	\begin{overpic}[width=\linewidth]{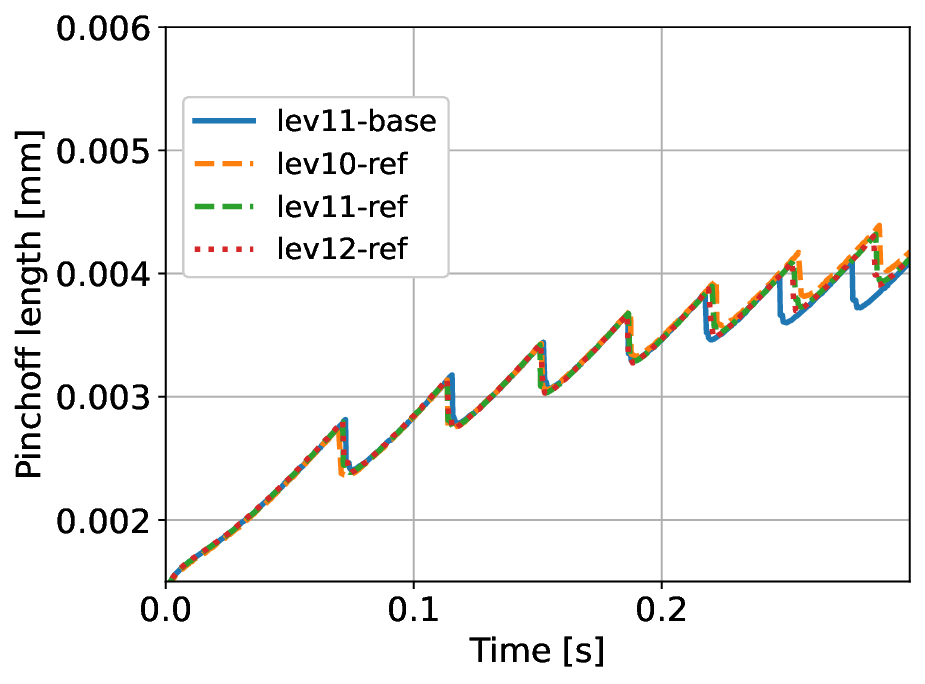}
		\put(5,74){\textbf{(b)}}
	\end{overpic}
\end{minipage}
\caption{Thread length: minima represents pinchoff position. (a): dripping case. (b): jetting case. Evaluation of different maximum level of refinement and different adaptivity threshold criteria.}
\label{Fig:pinchoff}
\end{figure}

For the sake of completeness, we show in Figure \ref{Fig:pinchoff} the time evolution of the jet length. Every jump corresponds to a pinch-off event and every minimum is the position of the jet immediately after pinch-off and {subsequent} retraction of the tip. As expected, in the dripping regime, Figure \ref{Fig:pinchoff}(a), the break-up position is stable, while in the jetting regime, Figure \ref{Fig:pinchoff}(b), it gradually drifts downstream, and even tiny deviations of the {pinch-off} position at each cycle of formation {accumulate and result} in unmatched positions with different grids at long times {when the adaptivity threshold is modified, while changing only the maximum level of refinement maintains an excellent convergence, see Figure \ref{Fig:pinchoff}(b). In the following, however, we will focus solely on the prediction of droplet formation regimes and diameter, and only the trend of the pinch-off position will be used to classify the regime, which remains consistent across all the tested grids.} 

\subsection{Regime maps} \label{sec:regimes}
\begin{figure}[h]
\center
\begin{minipage}[t]{0.48\textwidth}
	\centering
	\begin{overpic}[width=\linewidth]{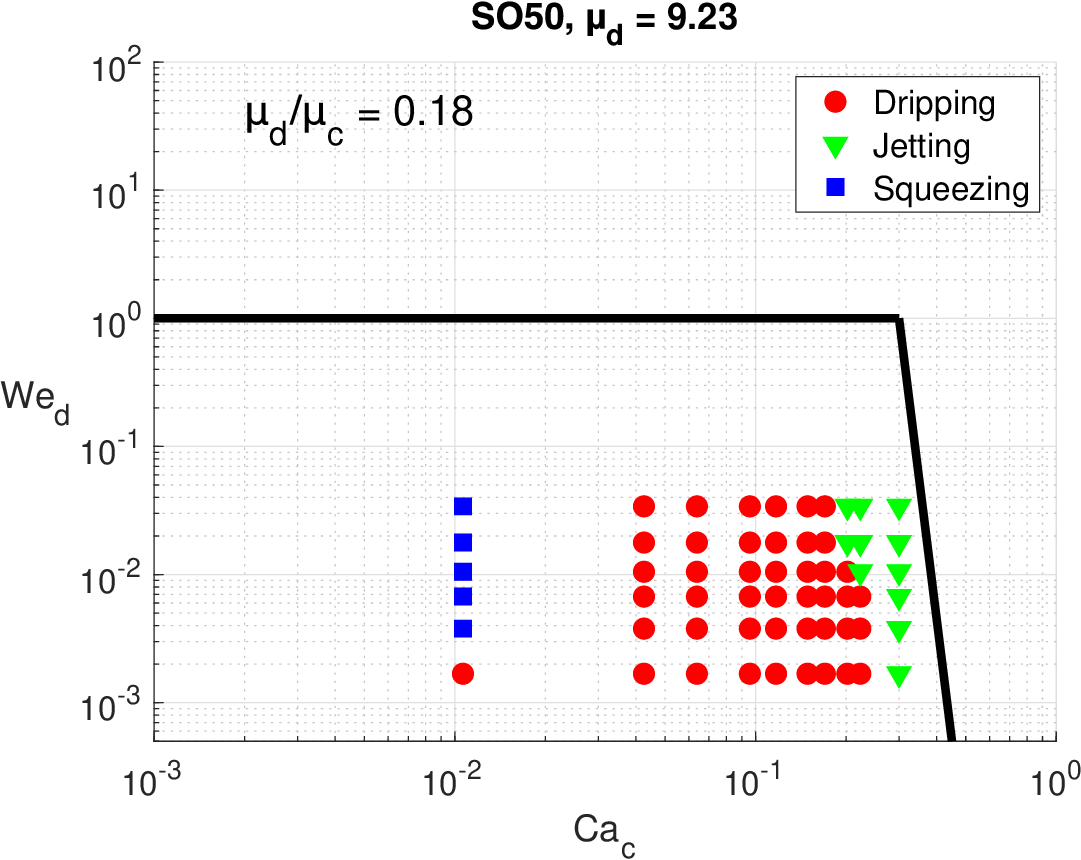}
		\put(5,77){\textbf{(a)}}
	\end{overpic}
\end{minipage}\hfill
\begin{minipage}[t]{0.48\textwidth}
	\centering
	\begin{overpic}[width=\linewidth]{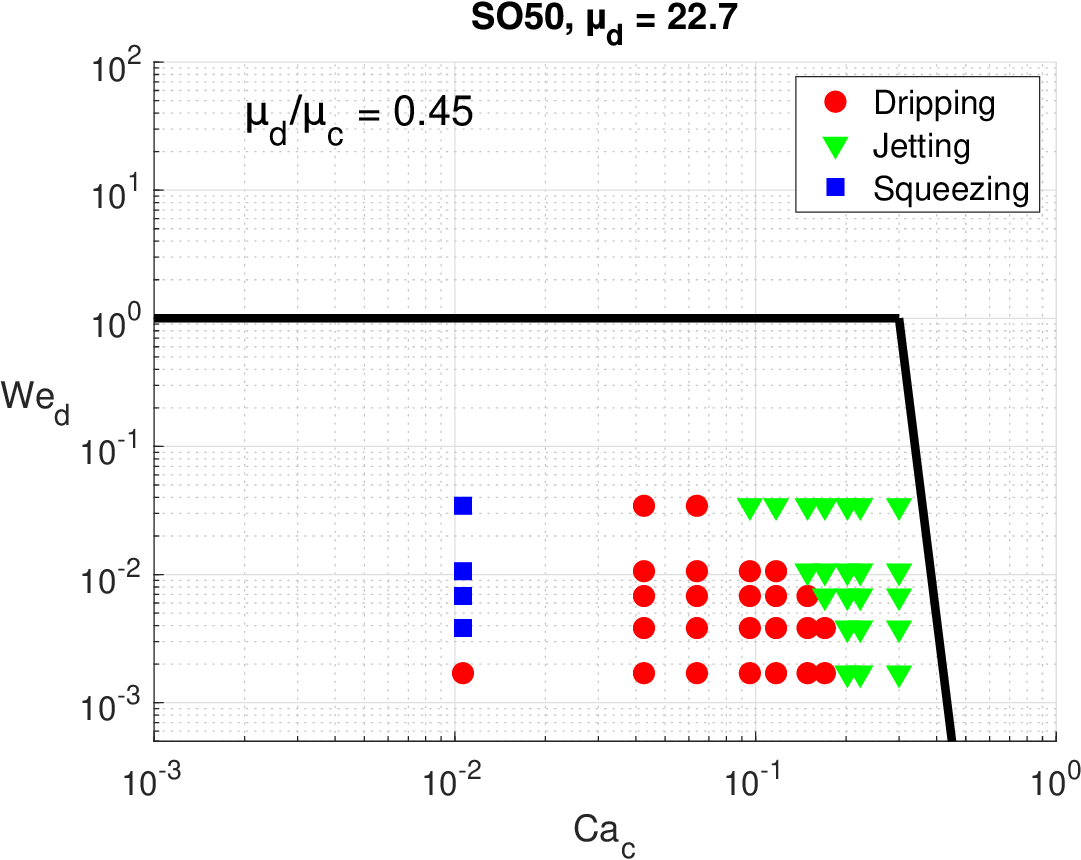}
		\put(5,77){\textbf{(b)}}
	\end{overpic}
\end{minipage}\hfill \\
\begin{minipage}[t]{0.48\textwidth}
	\centering
	\begin{overpic}[width=\linewidth]{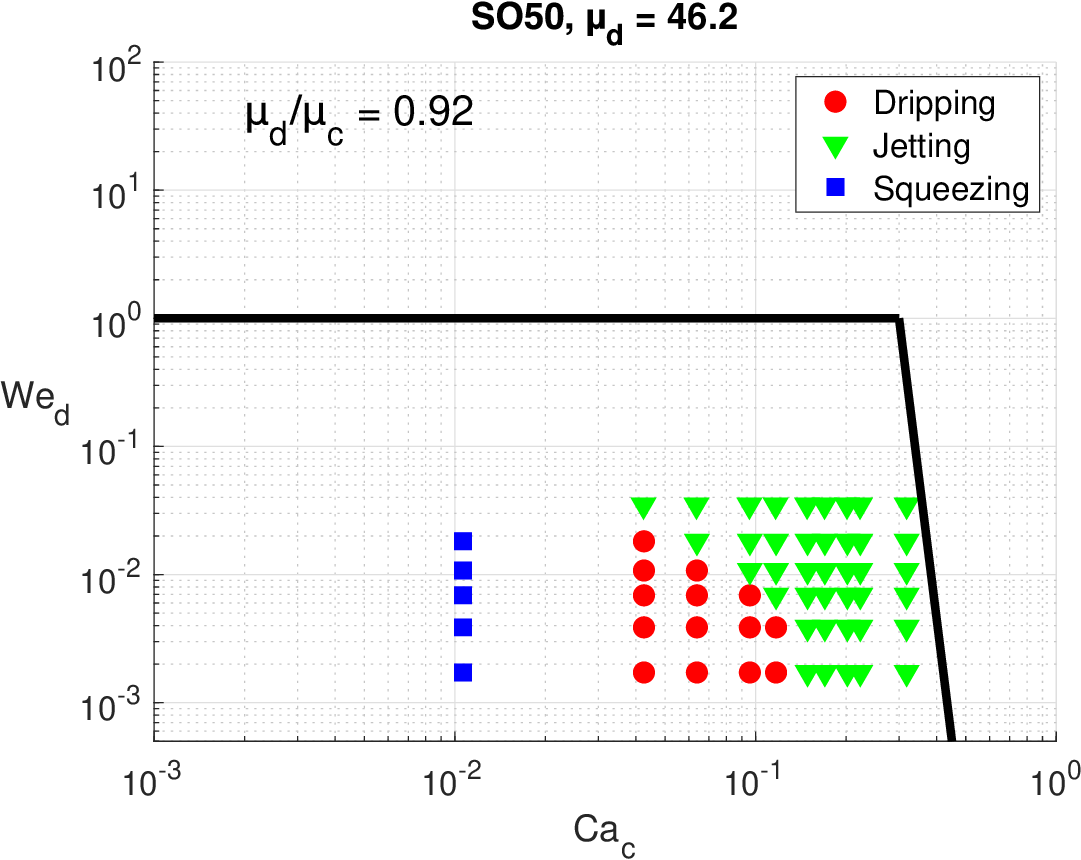}
		\put(5,77){\textbf{(c)}}
	\end{overpic}
\end{minipage}\hfill
\begin{minipage}[t]{0.48\textwidth}
	\centering
	\begin{overpic}[width=\linewidth]{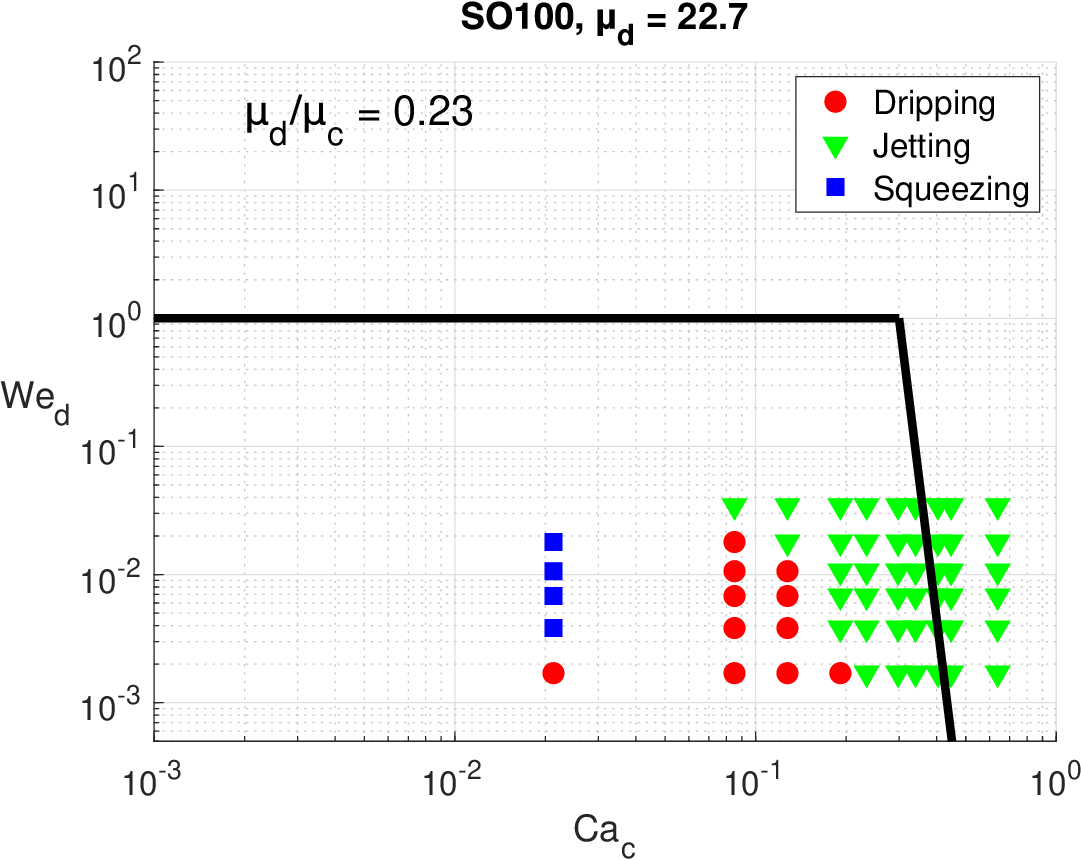}
		\put(5,77){\textbf{(d)}}
	\end{overpic}
\end{minipage}\hfill
\caption{Maps of droplet generation regimes as a function of $Ca_c$ and $We_d$. Blue squares {denote} squeezing, red circles dripping, and green triangles jetting. Black lines are the transition zones of Utada \textit{et al.} \cite{utada2007dripping} Fig. 4. Each panel represents a different {set} of simulations.}
\label{Fig:maps_weber}
\end{figure}

In this section we evaluate how the regime of droplet formation changes {with variations in} physical parameters and flow rates. We have performed a systematic parametric study structured in 4 {sets} of simulations, as reported below:
\begin{enumerate}[noitemsep]
\item continuous phase: SO50; dispersed phase PSF $6\%$ ($\mu_d = 9.23$);
\item continuous phase: SO50; dispersed phase PSF $8\%$ ($\mu_d = 22.7$);
\item continuous phase: SO50; dispersed phase PSF $10\%$ ($\mu_d = 46.2$);
\item continuous phase: SO100; dispersed phase PSF $8\%$ ($\mu_d = 22.7$);
\end{enumerate}
 {In this way, we can assess the influence of the dispersed-phase properties by comparing simulation sets 1–3, whereas simulation sets 2 and 4 are used to investigate the effect of the continuous-phase properties while keeping the dispersed-phase properties fixed.}
{For each set of physical parameters we have carried out a number of simulations between $50 - 60$, varying the inlet velocities of the two phases, with $u_c \in [1 ; 4 ; 6 ; 9 ; 11 ; \allowbreak 14 ; 16 ; 19 ; 21; 30] mm/s$, $u_d \in [4 ; 6 ; 8 ; 10 ; 13; 18] mm/s$.} In Figure \ref{Fig:maps_weber} we show the results of the 4 {sets} of simulations, in which we have identified the regimes of droplet generation in terms of {capillary}  number of the continuous phase and Weber number of the dispersed phase. We identify the different regimes with the following criteria: i) squeezing, when the interface reaches the wall of the duct and the volume of the formed droplet is greater than the maximum volume of a spherical droplet fitting the duct; ii) dripping when the pinch-off is stable in time and occurs always at the same position; iii) jetting, when the pinch-off position drifts downstream in time. 

\begin{figure}[h]
\center
{\includegraphics[width=0.6\textwidth]{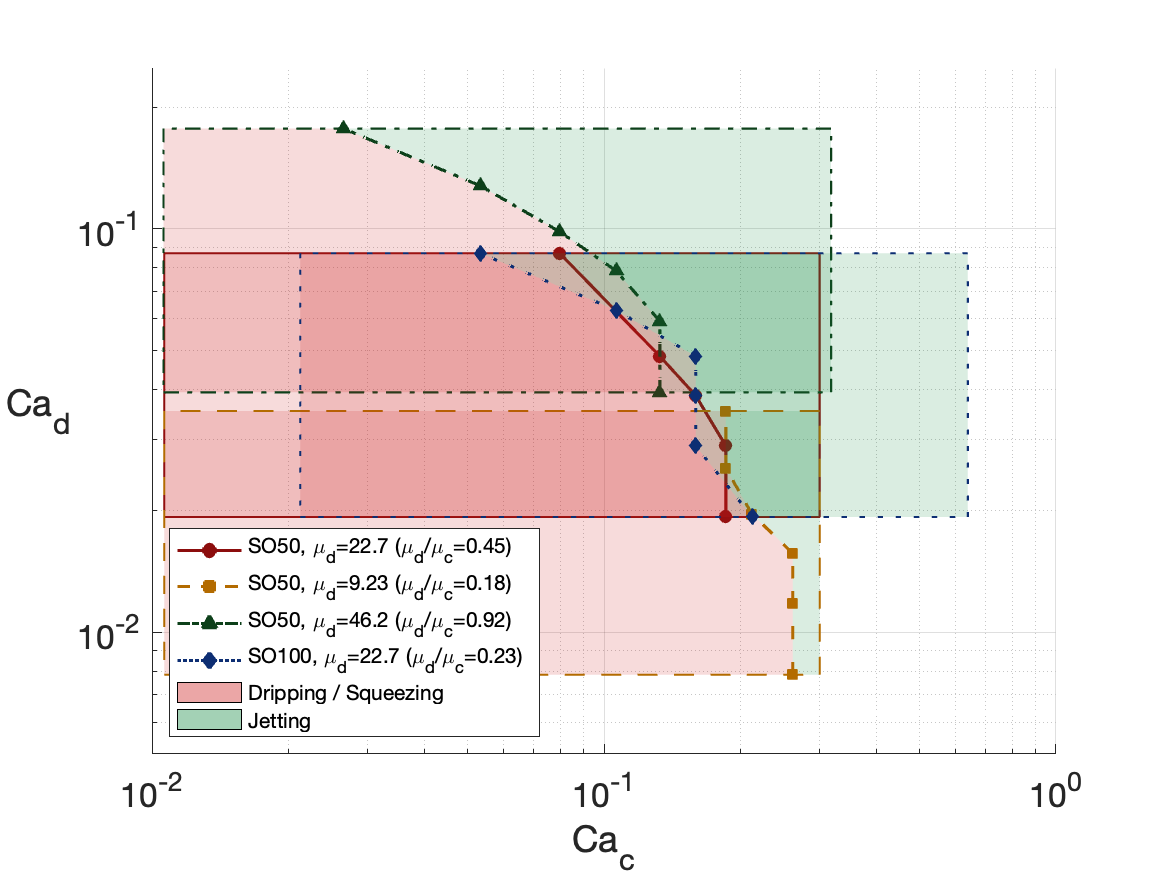}}
\caption{
{Regime maps as a function of the continuous-phase and dispersed-phase capillary numbers, $Ca_c$ and $Ca_d$, for the four sets of simulations. Each rectangle delimits the ($Ca_c$, $Ca_d$) range explored by one set of simulations and is split by that set's transition curve into a dripping/squeezing region (red) and a jetting region (green). Darker areas denote the overlap of two or more sets.}}
\label{Fig:maps_Ca}
\end{figure}

The black lines in Figure \ref{Fig:maps_weber} {denote} the transition zones found in the experimental work of Utada \textit{et al.} \cite{utada2007dripping}, namely jet thinning, at low Weber numbers, with increasing {capillary}  numbers, and jet thickening, at low {capillary}  numbers and increasing Weber numbers. It is worth noting that in \cite{utada2007dripping}, although they vary the viscosity ratio in the range $\mu_d / \mu_c = 0.01 - 10$, the jet thinning transition has mostly been studied for viscosity ratios in the lower end of the range. Our results focus on the jet thinning transition and show that the transition is strongly affected at least by the viscosity ratio of the two phases. The first noticeable result is that increasing the viscosity ratio, i.e. from $\mu_d/\mu_c = 0.18$ up to $\mu_d/\mu_c = 0.92$, {keeping fixed the continuous phase viscosity ($SO50$)}, {shifts} the transition {to lower} {capillary}  numbers. At the same time, we can also see from the comparison of {sets} 1 and 4, which are at similar viscosity ratios, that the transition zone is remarkably different. Keeping in mind the definition of $We_d = Re_d \cdot Ca_d$, and that the other dimensionless groups are fixed ($Ca_c$, $\mu_d/\mu_c$, $u_d/u_c$), it indicates that it is likely to be linked to the separate effect of $Re_d$ or $Ca_d$. Indeed, as we can see from Figure \ref{Fig:maps_Ca}, mapping the regimes in terms of the two {capillary}  numbers leads to a better overlap of the transition line. It {also }implicitly means that the dependence on $Re_d$ is {weak}. Results are in good qualitative agreement with the experimental map in \cite{lan2015numerical}.

{From a more physical point of view, increasing the continuous-phase capillary number $Ca_c$ means that the external shear increases relative to surface tension, resulting in smaller droplet diameters in the dripping regime, since the marginal stability condition is reached earlier, at smaller diameters.} The diameter reduces down to the tip diameter, then the flow transitions to jetting \cite{utada2007dripping}.

\subsection{Droplet diameters and scaling laws}
\begin{figure}[h]
\center
\begin{minipage}[t]{0.48\textwidth}
	\centering
	\begin{overpic}[width=\linewidth]{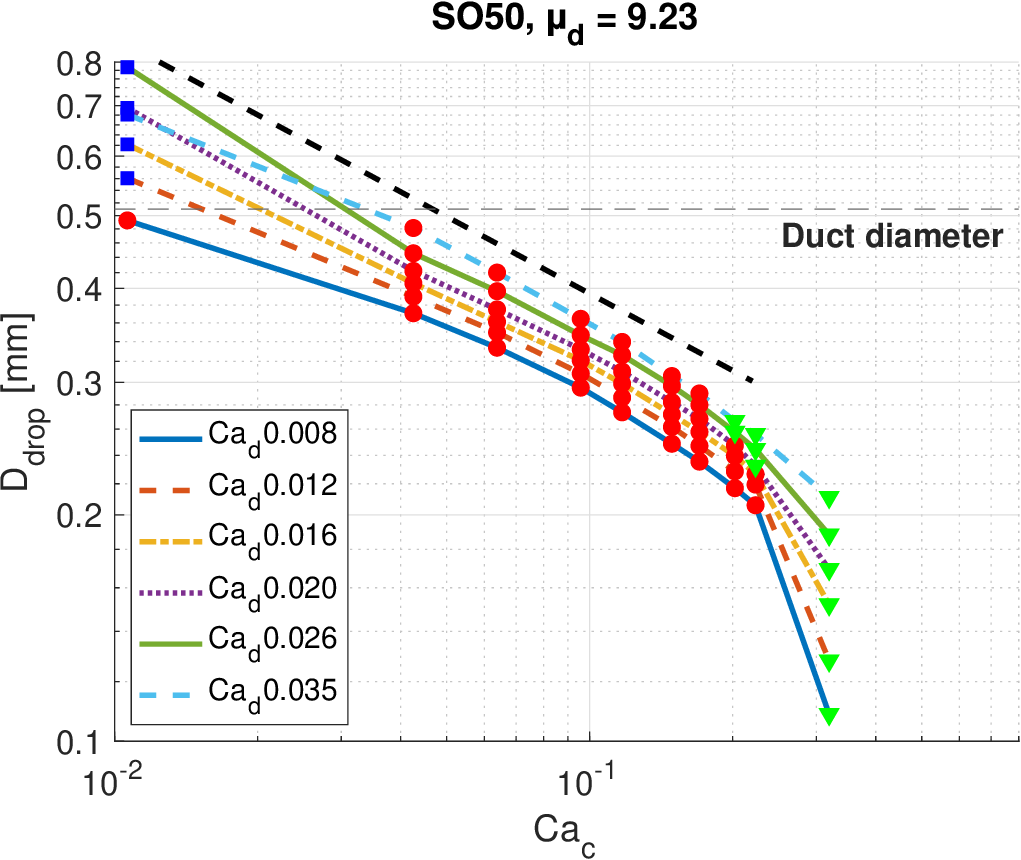}
		\put(5,82){\textbf{(a)}}
	\end{overpic}
\end{minipage}\hfill
\begin{minipage}[t]{0.48\textwidth}
	\centering
	\begin{overpic}[width=\linewidth]{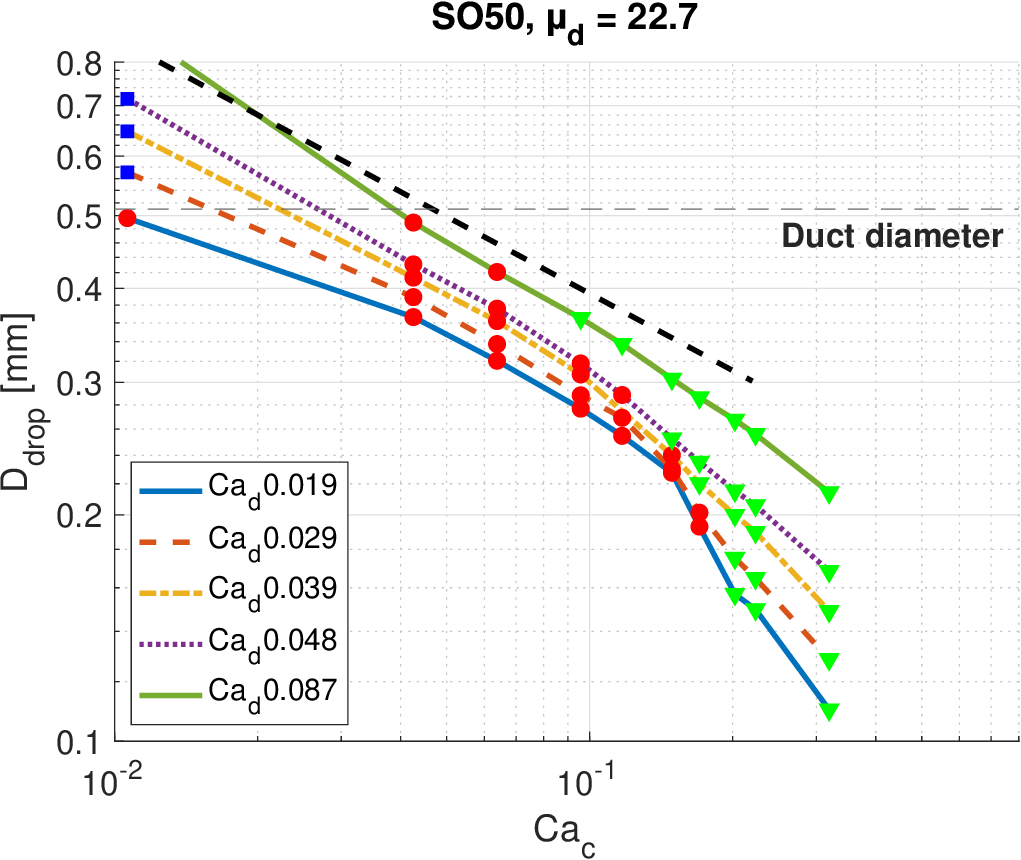}
		\put(5,82){\textbf{(b)}}
	\end{overpic}
\end{minipage}\hfill
\caption{Diameter of the generated droplets for the {sets} 1 (panel(a)) and 2 (panel(b)). Each line corresponds to a different value of $Ca_d$. Blue squares are for {droplets} in the squeezing regime, red circles for dripping and green triangles for jetting. The dashed black line indicates the slope of $-0.34$.}
\label{Fig:diameters}
\end{figure}

In this section we show how droplet diameters change with respect to the dimensionless numbers and the relative physical implications. The methodology for the evaluation of droplet diameters is the same {as that} used in Section \ref{sec:grid_convergence}. {As a fixed choice}, we have considered for all the simulations the diameter of the second generated droplet immediately after the pinch-off, {with the exception of the simulations at $u_c = 30 \, mm/s$, for which it is necessary to wait for the generation of the third droplet to reach a converged diameter}. In Figure \ref{Fig:diameters} we show the variations of the droplet diameters with respect to the {capillary}  numbers of the two phases for two different values of the dispersed phase viscosity (or of the viscosity ratio, since the viscosity of the continuous phase is fixed). First of all, from a qualitative point of view, we see that increasing the {capillary}  number of the continuous phase results in a decrease of the droplet diameter, which is straightforward since it is related to an increase of the viscous external forces on the droplet, that are destabilizing, with respect to surface tension, that is stabilizing. Therefore, the marginal stability condition occurs before, i.e. at smaller diameters of the tip.  

Variations of the {capillary}  number of the dispersed phase, keeping both $Ca_c$ and $\mu_d / \mu_c$ fixed, can be thought {of} as variations of the velocity ratio, i.e.:
\begin{equation}
Ca_d = \frac{\mu_d u_d}{\sigma} = \frac{\mu_c u_c}{\sigma} \frac{\mu_d}{\mu_c} \frac{u_d}{u_c} \, .
\end{equation}
Therefore, it is again physically consistent that increasing $Ca_d$ results in bigger droplet diameters, due to an increased flow rate of the dispersed phase that contributes to a greater supply during the growing phase. Finally, concerning the effect of viscosities, we see from Figure \ref{Fig:diameters} how increasing the viscosity ratio $\mu_d/\mu_c$ slightly decreases droplet diameters, which is due to an increased resistance to the passage of the dispersed phase through the neck. If we focus on the dripping regime (red circles in Figure), we can see that the scaling with respect to $Ca_c$ is consistent with the power-law $d_{drop} \propto Ca_c^{-0.34}$ proposed in \cite{lan2015numerical}. However, as stated above, a dependence on the dispersed phase {capillary}  number and on the viscosity ratio should also be taken into account. Moreover the scaling fits well at low {capillary} numbers, but deviates at the larger ones where the regime begins to transition to jetting.

\begin{figure}[h]
\center
\begin{minipage}[t]{0.48\textwidth}
	\centering
	\begin{overpic}[width=\linewidth]{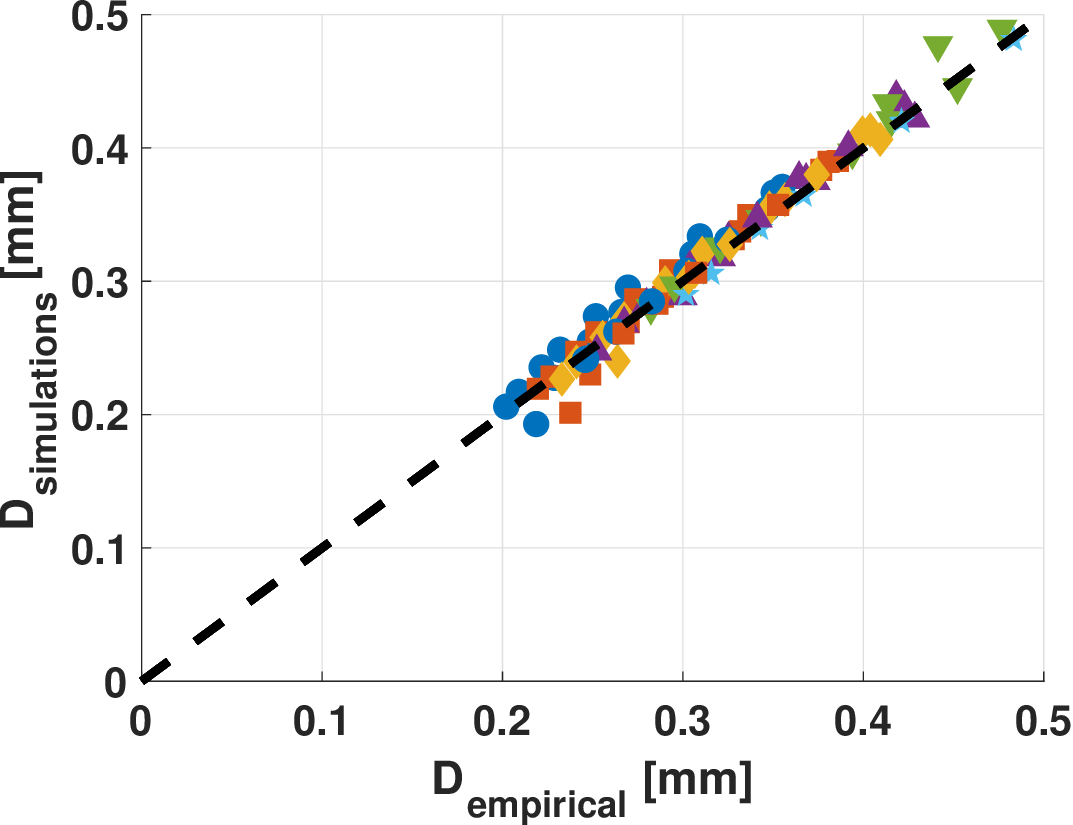}
		\put(15,70){\textbf{(a)}}
	\end{overpic}
\end{minipage}\hfill
\begin{minipage}[t]{0.48\textwidth}
	\centering
	\begin{overpic}[width=\linewidth]{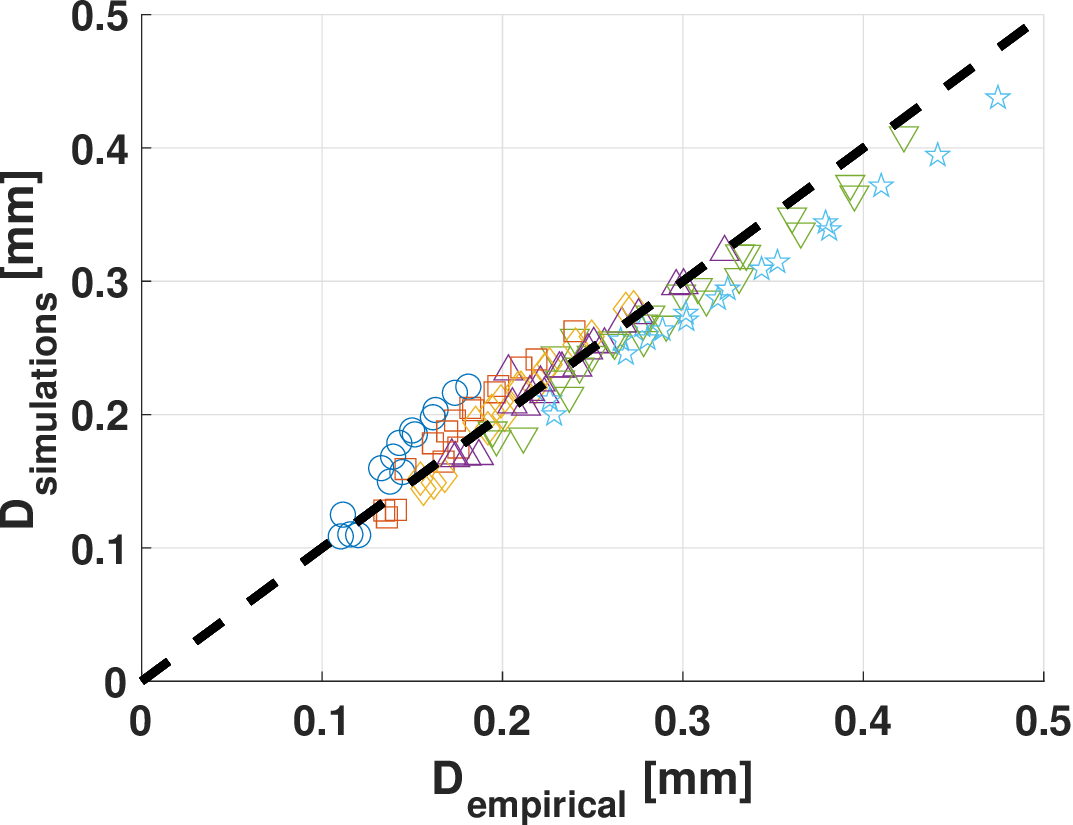}
		\put(15,70){\textbf{(b)}}
	\end{overpic}
\end{minipage}
\caption{Correlations between droplet diameters from simulations and from semi-empirical relations Eq. \ref{eq:relation_dripping}. (a): dripping regime; (b): jetting regime. Different symbols {correspond} to different values of $u_d$: $u_d = 4$ (blue circle), $u_d=6$ (red square), $u_d=8$ (yellow diamond), $u_d=10$ (upward purple triangle), $u_d=13$(downward green triangle), $u_d=18$(light blue star).}
\label{Fig:correlations}
\end{figure}

Based on these {observations,} we {looked for} scaling laws that could be used to accurately predict droplet diameters in the dripping and jetting regimes, that, if combined with the regime maps of section \ref{sec:regimes}, can help {in} devising efficient microfluidic devices. In the dripping regime Lan \textit{et al.} \cite{lan2015numerical} have proposed a scaling law based on the continuous phase {capillary}  number and the dispersed phase Reynolds number, i.e. $d_{drop}/d_{in} \propto Ca_c^{-0.34} Re_d^{0.06}$. Indeed, as pointed out in the analysis of the regime maps in terms of Weber number or {capillary}  number, we expect a small effect of $Re_d$ on the dynamics and on the size of the droplets, and this is confirmed by the small exponent that they found in the above relation based on their experimental results. Starting from this consideration, and following the criteria found in the analysis of regime maps and droplet diameters, we {sought} a dependence of the droplet diameter of this kind: $d_{drop}/D = f(Ca_c, Ca_d, \mu_d/\mu_c, Re_d)$. {We performed a linear regression using all the simulations from all the sets belonging to the dripping regime to determine the power-law exponents, obtaining the following expression:}
{
\begin{equation} \label{eq:relation_dripping}
\frac{d_{drop}}{D} = 0.226 \cdot Ca_c^{-0.34} \cdot Ca_d^{0.21} \cdot \Bigl ( \frac{\mu_d}{\mu_c} \Bigl)^{-0.23} \cdot Re_d^{-0.005} \, . 
\end{equation}
}
The dependence on $Ca_c$ and the negligible effect of $Re_d$ are in very close agreement with \cite{lan2015numerical}, while the additional dependence on $Ca_d$ and $\mu_d/\mu_c$ {shows} significant exponents. Figure \ref{Fig:correlations}(a) shows the correlations between simulations and the above semi-empirical relation with an $R^2$ correlation coefficient of $R^2=0.96$.

Concerning the scaling for the jetting regime, we started from the theoretical arguments and experimental {observations} of Utada \textit{et al.} \cite{utada2007dripping} and Lan \textit{et al.} \cite{lan2015numerical} that {support} that within the Rayleigh-Plateau instability the droplet diameter is proportional to the diameter of the jet, which is due to the fact that the most unstable mode wavelength is proportional to the jet diameter, and that the volume of the forming droplet can be considered equal to the volume contained in one wavelength \cite{utada2007dripping}, therefore $d_{drop} \simeq 2 d_{jet}$. At the same time, under the assumption of a Stokes flow, the diameter of a co-flowing jet can be expressed in terms of the flow rate ratio $2 (d_{jet} / D)^2 \simeq Q_d / (Q_c + Q_d)$, at leading order. Based on this theoretical relation, we {seek} a dependence of this kind with an additional effect of the viscosity ratio, as is evident from the above analysis. The result of a linear regression with imposed dependence on the flow rate ratio, i.e. imposing $d_{drop}/D  = C \cdot  [2Q_d / (Q_c + Q_d)]^{0.5} \cdot (\mu_d/\mu_c)^{\alpha}$, yields:
{
\begin{equation}
d_{drop}/D  = 2.34 \cdot  \Bigl ( \frac{Q_d}{(Q_c + Q_d)/2} \Bigl )^{0.5} \cdot \Bigl (\frac{\mu_d}{\mu_c} \Bigl)^{0.052} \, .
\end{equation}
}
Figure \ref{Fig:correlations}(b) shows the correlations between {the simulation results} and the above relation, which gives an $R^2$ correlation coefficient of {$R^2 = 0.92$}.

\section{Discussion and Conclusions}
In this work we have analyzed the suitability of the open-source software Basilisk to reproduce microdroplet generation mechanisms in {micro- and millifluidic devices}. We have assessed the accuracy and the usability of Basilisk starting from the analysis of the Taylor bubble configuration, where we have found excellent agreement with theoretical and empirical literature laws for droplet terminal velocity, analyzing also the sensitivity to the main numerical parameters. {The present work also provides useful insights into the choice of the refinement threshold and other numerical parameters.}

We have then focused on axially-symmetric configurations, first of all {reproducing} the experimental results of Lan \textit{et al.} \cite{lan2015numerical} in the two notable regimes of dripping and jetting. Secondly, we evaluated how physical parameters {affect} the regimes of droplet {formation} and the associated diameters, via {$4$ distinct sets of simulations}, each one with fixed viscosity and density of the two phases, varying the inlet velocities. In such {a} way it has been possible to map the regimes of droplet generation, at first in terms of $Ca_c$ and $We_d$, then more effectively in terms of $Ca_c$ and $Ca_d$, finding a distinct separation zone from the dripping to the jetting regime that {overlaps} consistently for the $4$ {sets} of simulations. For {the simulations in the dripping and jetting regimes, we} have also analyzed in {detail }the size of the generated droplets and how they vary with respect to physical parameters, quantifying the decrease with increasing continuous phase {capillary}  number and with increasing dispersed phase {capillary}  number, in the dripping regime, and assessing the relevant dependence on the viscosity ratio in both regimes. Semi-empirical scaling laws have then been proposed for the two regimes, based on these new findings and inspired {by} previous literature works \cite{lan2015numerical, utada2007dripping}. 

Results on regime maps, obtained from a systematic study, support and complement literature results and can be of valuable importance for {the} preliminary design of microfluidic coaxial devices. Similar analysis can then be extended to different layouts and configurations. Moreover, the semi-empirical scaling laws of droplet diameters can be {suitably} used in the search for an optimal balance between production rates and droplet size. {Beyond providing quantitative results and laws that may be of practical relevance,} as explained above, we have further investigated the physical reasons behind these mechanisms, emphasizing the importance of taking the viscosity ratio into account.

This work also paves the way for the use of Basilisk software in this type of microfluidic applications, thus making it possible to take advantage of its high accuracy, resulting from the sharp treatment of the interfaces, together with the benefits provided by the use of adaptive grids. Fully 3-dimensional simulations, also with more complex geometries, that will overcome the {current} limitation of the combined use of the embedded framework with contact angle dynamics only in {two} dimensions, will soon be possible thanks to the continuous development and improvement of the existing functions and solvers (see Tavares \textit{et al.} \cite{tavares2024coupled} for recent developments on contact angle dynamics).

\section*{Acknowledgments}

This work is supported by PNRR M4C2 - HPC, Big data and Quantum Computing (Simulazioni, calcolo e analisi dei dati e altre prestazioni - CN1) - CUP I53C22000690001 SPOKE 6 Multiscale modeling \& Engineering applications.
We also acknowledge the CINECA award under the ISCRA initiative for the availability of high performance computing resources and support.

\section*{Declaration of generative AI and AI-assisted technologies in the manuscript preparation process}
During the preparation of this work, the author(s) used ChatGPT for English editing. The author(s) reviewed and edited the output as needed and take full responsibility for the content of the published article. 

\newpage










  
\bibliographystyle{unsrt}
\bibliography{bibliografia}

\end{document}